\documentclass{aastex}

\shorttitle{RSGs in the Magellanic Clouds}
\shortauthors{Massey \& Olsen}

\begin{document}

\title{The Evolution of Massive Stars. I. Red Supergiants in the 
Magellanic Clouds}

\author{Philip Massey\altaffilmark{1}}

\affil{Lowell Observatory, 1400 W. Mars Hill Road, Flagstaff, AZ 86001}

\email{Phil.Massey@lowell.edu}

\and

\author{K. A. G. Olsen\altaffilmark{1}}
\affil{Cerro Tololo Inter-American Observatory, National Optical Astronomy
Observatory, Casilla 603, La Serena, Chile}
\email{kolsen@noao.edu}

\altaffiltext{1}{Visiting Astronomer, Cerro Tololo Inter-American Observatory,
National Optical Astronomy Observatory, which is
operated by the Association of Universities for Research in Astronomy,
Inc., under cooperative agreement with the National Science Foundation.}

\begin{abstract}

We investigate the red supergiant (RSG) content of the SMC and LMC using 
multi-object spectroscopy on a sample of red stars previously identified
by {\it BVR} CCD photometry.
We obtained high accuracy ($<1$ km s$^{-1}$) radial velocities for
118 red stars seen towards the SMC and 167 red stars seen towards the
LMC, confirming most of these (89\% and 95\%, respectively) as 
red supergiants (RSGs).  
Spectral types were also determined for most of  
these RSGs.  We find that the distribution of spectral types is
skewed towards earlier type at lower metallicities: 
the average (median) spectral type is K5-7~I in the SMC,
M1~I in the LMC, and M2~I in the Milky Way. 
Our examination of the Kurucz Atlas 9 model atmospheres suggests that the effect
that metallicity has on the appearance on the TiO lines is probably
sufficient to account for this effect, and we argue that RSGs in the
Magellanic Clouds are 100$^\circ$K (LMC) and 300$^\circ$K (SMC) cooler
than
Galactic stars of the same spectral types. The
colors of the Kurucz models are not consistent
with this interpretation for the SMC, although other models 
(e.g., Bessel et al.) show good agreement.  A finer grid of higher-resolution
synthetic spectra appropriate to cool supergiants is needed to better
determine the effective temperature scale. We
compare the distribution of RSGs in the H-R diagram to that of various
stellar evolutionary models; we find that none of the models produce 
RSGs as cool and luminous as what is actually observed. This result is
much larger than any uncertainty in the effective temperature scale.
We note that were we to simply
adopt the uncorrected
Galactic effective scale for RSGs and apply this to our sample, then the
SMC's RSGs would be under luminous compared to the LMC's, contrary to what
we expect from stellar evolution considerations.  In all of our H-R diagrams,
however, there is an elegant sequence of decreasing effective temperatures
with increasing luminosities; explaining this will be an important test
of future stellar evolutionary models.
Finally, we compute the blue-to-red supergiant ratio in the SMC and LMC,
finding that the values are indistinguishable ($\sim 15$) for the two
Clouds.  We emphasize that ``observed" B/R values must be carefully
determined if a comparison to that predicted by
stellar models is to be meaningful.  The non-rotation Geneva models 
overestimate the number of blue-to-red supergiants for the SMC, but
underestimate it for the LMC; however, given the inability to produce 
high luminosity RSGs in the models that match what is observed in the
H-R diagram, such a disagreement is not surprising. 

\end{abstract}


\section{Introduction}

The evolution of massive stars will depend upon the initial metallicity
of the gas out of which they form, and thus we can expect 
differences in the relative numbers of various stages of massive stars
among nearby galaxies. (For a comprehensive review of the subject,
see Maeder \& Conti 1994.)  The primary effect that metallicity has is due
to its influence on radiatively-driven stellar winds and the resulting
mass loss.  Typical mass-loss rates for Galactic O-type stars are
0.5--20$\times 10^{-6} \cal M_\odot$ year$^{-1}$ (Puls et al.\ 1996),
with the more massive stars losing a greater fraction of their mass
during their main-sequence lifetimes\footnote{Since $\dot{M}$ 
depends upon the luminosity $L$
as $\dot{M}\sim L^{1.7}$ (Pauldrach, Puls \& Kudritzki 1986;
de Jager, Nieuwenhuijzen, \& van der Hucht 1988; 
Kudritzki \& Puls 2000), and since
luminosity depends upon mass $M$ as $L\sim M^2$ for high mass
stars (Massey 1998a,
using the Schaller et al.\ 1992 $Z=0.02$ evolutionary tracks), we
expect that the mass-loss rates will depend upon the mass roughly as
$\dot{M}\sim M^{3.4}$.  The main-sequence lifetime $\tau{\rm}$ is a
relatively weak function of the mass for high-mass stars, and inspection
of the Schaller et al.~(1992) $Z=0.02$ tracks suggests that
$\tau_{\rm ms}\sim M^{-0.6}$. 
So we expect that the total mass loss during the 
main-sequence phase ($\Delta M= \dot{M}\times \tau_{\rm ms}$) will 
go roughly as $\Delta M \sim M^{2.8}$.  Thus the fractional mass lost,
$\Delta M/M$, will go as $M^{1.8}$.  And, this is just on the main-sequence!
Stars with luminosities above $\log L/L_\odot \sim 5.8$ will suffer
episodes of enhanced mass loss as their luminosities exceed the
Eddington limit once line opacities are taken into account; this
stage is likely identified with the Luminous Blue Variable phase
(stars such as $\eta$ Car and S Dor), and accounts for the
Humphreys \& Davidson (1979) upper luminosity limit in the H-R diagram
(Lamers 1997).}.  A very high mass star (100$\cal M_\odot$) might then
lose 50\% of its mass during its evolution, which has a profound effect
on its path in the H-R diagram, as first
shown by 
de Loore, De Gr\`{e}ve, \& Lamers (1977), 
de Loore, De Gr\`{e}ve, \& Vanbeveren (1978), 
Chiosi, Nasi, \& Sreenivasan (1978),  
Chiosi, Nasi, \& Bertelli (1979), 
Brunish \& Truran (1982), 
and subsequent
investigations. Mass-loss rates will scale with metallicity $Z$ to some
power, with the exponent variously estimated from 1.0 to 0.5 
(Abbott 1982;
Lamers \& Cassinelli 1996; Kudritzki et al.~1989; 
Puls, Springmann, \& Lennon 2000; Kudritzki \& Puls 2000;
Vink, de Koter, \& Lamers 2001; 
Kudritzki 2002).  
Beyond the main-sequence mass-loss
rates are highly uncertain; for instance, mass loss during the LBV phase
is highly episodic and large, with little agreement in what
drives the outbursts (Humphreys \& Davidson 1994, Maeder \& Conti 1994).
Large uncertainties also exist in the mass-loss rates during the red
supergiant (RSG) phase, making the subsequent tracks even less certain
(Salasnich, Bressan, \& Chiosi 1999).  It is commonly assumed that mass-loss
rates for Wolf-Rayet stars (WRs) are independent of initial metallicity, since
their atmospheres have been so enriched by the products of their own
nuclear burning (e.g., Schaller et al.~1992), but Crowther et al.~(2002)
have recently argued that iron is an important element in driving the WR
winds, and hence that there will be 
some $Z$ dependence in their mass-loss
rates.

In addition to the effects of mass loss, stellar evolutionary tracks are
also sensitive to the treatments of convection and mixing
(Maeder \& Meynet 1987), and there is considerable disagreement
among the pundits as to the proper way to include these in
the models (Maeder \& Conti 1994).  Recent
emphasis has been on the role that rotation plays in 
mixing in massive stars (Maeder \& Meynet 2000, 2002).  
Convection
and mixing also show some dependence on the metallicity
(see, for example, Meynet \& Maeder 2002), and the uncertainties
in their treatment underscores the fact that the physics of massive
star evolution is not perfectly well understood at present.

In order to advance our understanding of massive star evolution, it is
necessary to have a solid observational database with which the predictions
of stellar evolutionary theory may be compared and refined.   A well-known example is the relative number of blue and red supergiants, which van den Bergh (1973) first suggested varied among nearby galaxies due to the effects of metallicity on massive star evolution. Particularly
sensitive tests include the relative numbers of different types of evolved
massive stars, such as the relative number of different types of Wolf-Rayet
stars (WC-type and WN-type), or the relative number of RSGs and WRs.
Maeder, Lequeux, \& Azzopardi (1980) proposed that the latter number ratio
would be particularly sensitive to metallicity effects.  

However, there are many observational difficulties 
in determining such statistics
reliably.  For Wolf-Rayet stars, there are selection effects against finding
WN-type WRs (Armandroff \& Massey 1985, Massey \& Johnson 1998).  For
red supergiants, the problem is that when we look towards a galaxy such
as M~31 or the Magellanic Clouds we see not only the {\it bona fide} 
extragalactic RSGs but also foreground galactic red dwarfs in the same
color and apparent magnitude range. Massey (1998b) found that {\it BVR}
photometry helped separate RSGs from foreground dwarfs, but was not by
itself sufficient.  Spectroscopy allows an accurate assessment, however.
Although the luminosity indicators for late-type stars are rather 
subtle, an effective technique is to use the near-IR Ca~II triplet lines
 to determine a star's radial velocity.
For many Local Group galaxies this provides a very clean separation of
foreground red dwarfs from extragalactic red supergiants.

Here we extend this technique to our nearest galactic neighbors, the
Magellanic Clouds (MCs).  Massey (2002a) estimated the degree of
foreground contamination would be about 10\% in the appropriate
magnitude/color range, far lower than the $\sim$ 50\% found in M31,
M33, and NGC~6822 by Massey (1998b), both because the Clouds are nearer
and at higher Galactic latitude. However, an accurate census of the RSG
population in the Magellanic Clouds is of particular interest, as
these galaxies are sufficiently close that a great deal is already known
about their blue supergiant population, for which much spectroscopy
has been carried out (Massey et al.~1995, Massey 2002a).

Throughout this paper we will adopt the distance and average
reddenings listed by
van den Bergh (2000), namely $(m-M)_o=18.50$ and $E(B-V)$=0.13 for the LMC,
and $(m-M)_o=18.85$ and $E(B-V)=0.06$ for the SMC. 

\section{Observations and Reductions}

Our sample of red supergiant candidates comes from 
a recent {\it UBVR} CCD survey covering most of the Clouds (Massey 2002a).
The sample was chosen to include
potential K- and M-type supergiants, based upon the criteria
of $(V-R)_o>0.6$ and a $V$ cutoff such that $M_{\rm bol}<-7.0$ given
the observed $V-R$ color and assumed average reddening (i.e., Tables 9A
and 9B of Massey 2002a).  A few additional
red stars which did not quite meet these requirements
were also included.

Our spectroscopy used the Hydra fiber positioner (Barden \& Ingerson 1998)
on the Blanco 4-m
telescope at Cerro Tololo Inter-American Observatory during the nights
of (UT) 4-6 Oct 2001.  On the first two nights, grating 380 (1200 lines mm$^{-1}$, blaze 8000\AA) was used in first order with an RG-610 blocking 
filter to obtain data at the Ca~II triplet lines
($\lambda \lambda$ 8498, 8542, 8662). A SITe 2K $\times$ 4K 
CCD was used unbinned,
behind a 400~mm focal length camera on a bench spectrograph,
to obtain a dispersion of 0.27\AA\ pixel$^{-1}$, with a wavelength
coverage extending from
8000\AA\ to 9000\AA.  A 200 $\mu$m slit plate was inserted at the output
of the fiber bundle to yield a resolution of 1.2\AA\ (4.5 pixels).  
On
the third night (and for a small portion of an engineering night that
immediately preceded the run), we used grating KPGL1 (632 lines mm$^{-1}$,
blaze 4200\AA) in first order with no blocking filter to obtain
spectra in the blue in order to determine spectral subtypes.  The CCD
was binned in the spectral direction by a factor of 2, to obtain
a dispersion of 1.2\AA\ pixel$^{-1}$, with a wavelength coverage extending
from 3900\AA\ to 6100\AA.  No slit plate was used, and the resolution
(set by the size of the fibers) was approximately 4\AA\ (3.5 binned pixels).

The Hydra fiber positioner consists of 138 fibers (300$\mu$m, or 2.0-arcsec
in diameter) that can be accurately positioned in a 40-arcminute diameter
field of view at the RC focus of the Blanco 4-m. An atmospheric
dispersion corrector is mounted above the focal plane.  The closest
fiber-to-fiber spacing is approximately 25 arcsecs. This proved a good
match to the density of RSG candidates in most of our fields.  

Our observing procedure was to configure Hydra at the zenith, and then obtain
a short exposure of a quartz-lamp projector flat 
that could be used for flat fielding 
and for removing the relative transmissions of each fiber. (The projector
flats were taken for each new configuration to guard against slight 
flexure changes in the CCD dewar as the liquid nitrogen cryogen
evaporates.) We would then
offset to the field position and align the telescope using 3 to 7 ``field 
orientation probes" (bundles of 6 closely spaced fibers) that had been
placed at the coordinates of bright stars within the field; these would also
be used for guiding. Our program observations then consisted of 3 exposures
of 5 mins in length.  These would be followed by a short exposure of a 
comparison lamp of He, Ne, and Ar for wavelength calibration.  We would
then return to zenith and reconfigure for the next field.  The observations
were all carried out by K.A.G.O., while P.M. kibitzed from his office
in Flagstaff using the internet to help examine the data in real time.

Conditions were relatively good throughout the run, with  two hours
lost at the beginning of the first night due to fog and the last hour
of the third night lost to clouds.  The seeing was poor on the first
night (3 arcsecs) but was significantly better (1-2 arcsecs)
on subsequent nights.  The variation in throughput caused by the
changes in seeing have no effect on our results, as we are concerned
only with the relative strengths and positions of absorption features,
and not on absolute spectrophotometry.

All told, we were able to obtain radial velocity information for 6 fields
in the SMC and 10 fields in the LMC, with a repeat of one of the SMC fields
on the second night as a consistency check.  The same 6 SMC fields
were observed for the purposes of spectral classification, along with 7
of the LMC fields. 

On our two radial velocity nights we obtained a total
of 10 observations (5 per night)
of 4 radial velocity standards, spread throughout the night.
Several different fibers were used for the standards, and
the stars were chosen from
the list of standard radial velocity (RV) stars in
the 2001 {\it Astronomical Almanac}, selected for being of late-type
and accessible during our run.  The stars included 
HD~12029 (K2~III, RV=+38.6  km~s$^{-1}$),
HD~80170 (K5~III-IV, RV=0.0  km~s$^{-1}$),
HD~213947 (K2, RV=+16.7 km~s$^{-1}$), and
HD~223311 (K4~III, RV=$-$20.4 km~s$^{-1}$).
As we will describe in Section~\ref{sec:rvs} there was no systematic difference
from one night to the next, and our precision was sufficiently high to 
detect small inconsistencies in the relative velocities of the standards.

For the purposes of spectral classification, spectral standards were taken
from the list of Morgan \& Keenan (1973), and included
HD~160371 (K2.5~Ib), 
HD~52005 (K3~Ib),
HD~52877 (K7~Ib),
HD~42475 (M0-M1~Ib),
HD~42543 (M1-M2 Ia-Ib),
HD~36389 (M2~Iab-Ib),
HD~190788 (M3- Ib), and
HD~89845 (M4.5~Ia).

After basic CCD processing (overscan bias subtraction and trimming of the data),
we reduced the spectra using the IRAF ``dohydra" script. The quartz-lamp
projector flats were used to define the identification,
location, and shape of the fiber profiles
on the chip. This information was used to ``optimally extract" the program
objects and comparison exposures; flat-fielding and removing the fiber-to-fiber
variations was done using the extracted projector flat exposures as well.
We found that the illumination of the outlying fibers with the projector flat
did not match the sky illumination very well.  In sky-limited applications
this would compromise the sky subtraction unless corrected for by 
observations of blank sky, say, but since our stars were quite bright
compared to the sky, this made little difference in our final data.
The extracted spectra were then wavelength calibrated using the extracted
comparison-line spectra.
Finally, the three one-dimension spectra of each object were averaged using
bad-pixel rejection.  The standard star data were treated identically,
except that a single exposure was involved and so no averaging was done.
The spectra in the red (that would be used for radial velocity measurements)
were then normalized by a low-order cubic spline, and then shifted by unity 
to make the average continuum level zero.

\section{Analysis}

\subsection{Radial Velocities}
\label{sec:rvs}

Radial velocities were measured by cross-correlating each Magellanic Cloud
spectrum against each of the radial velocity standard star observations.
Since we could find no systematic effect in
cross-correlating the spectra of the radial velocity standards from one night
to the next, we simply treated all of our data the same, regardless of
which night they were obtained on.  We used the IRAF routine "fxcor",
and limited the cross correlation to the region 8450\AA-8700\AA\ in order
to isolate the Ca~II triplet ($\lambda \lambda$ 8498, 8542, 8662).
The cross-correlation peaks were fit by a parabola, resulting in an
internal precision of 0.5-0.7 km~s$^{-1}$ for each measurement.
The measurement based on each of the ten standard star observations were
then averaged; the agreement between these were excellent, and the 
resulting means had a standard deviation of the mean of 0.2-0.3 km~s$^{-1}$.
We believe this is an honest estimate of our actual accuracy, as quite
a few stars were observed twice (or even three times) owing either to their
locations in overlapping fields, or due to two observation of
the same fields on different nights.
The typical agreement for these stars was 0.25 km~s$^{-1}$.
Our spectra are so well exposed, and the Ca~II triplet lines so strong, that
we could easily detect small systematic differences in the cross-correlations
produced by different standard stars.  For instance, each of the two
observations of the standard star HD~213947 (obtained on separate nights)
produced cross-correlations that were $\sim 1$ km~s$^{-1}$ high compared to
that obtained from the ensemble, while the standard star HD~12029
produced cross-correlations that were $\sim 1$ km~s$^{-1}$ low compared to
that obtained from the ensemble\footnote{Specifically, if we adopt
the velocities of HD~213947 (16.7 km~s$^{-1}$) and HD~223311 ($-20.4$ km~s$^{-1}$) as correct, then the true radial velocity of HD~213947 is 15.0
km~s$^{-1}$ rather than the 16.7 km~s$^{-1}$ adopted by the IAU,
while that of HD~12029 is 39.6 km~s$^{-1}$ rather than the 38.6 km~s$^{-1}$
adopted by the IAU.}.

Altogether, we obtained radial velocities for 118 stars in the SMC.
Three were measured three times, and 42 were measured twice.
For the LMC, we obtained radial velocities for 167 stars.  Of
these, seven were measured three times, and 35 were measured twice.

In Tables~\ref{tab:SMC} and \ref{tab:LMC} 
we give the average radial velocities of the stars
in our sample.  Figure~\ref{fig:rvhists} 
shows the histograms of these velocities,
with the center-of-mass systemic velocities of the Clouds indicated.  

For both the SMC and LMC there is excellent agreement in the peaks of
the histograms and the cataloged systemic velocities of each Cloud.
The ``tail" of velocities extending to lower radial velocities is readily
identified as the foreground red dwarfs which we had hoped to distinguish
from the members of the Clouds.  For the LMC diagram the separation is
quite clean.  The lower systemic velocities of the SMC results in there
being a little uncertainty for three stars with intermediate velocities.
We have assigned membership in the tables based upon whether or not the
radial velocity is greater than 100 km~s$^{-1}$.  

Based upon the radial velocities, we conclude that 11.0\% of the stars in
the SMC sample proved to be foreground stars, while only 5.3\% of the stars in
the LMC sample were foreground stars. 

\subsection{Spectral Classification}

We include our spectral types in Tables~\ref{tab:SMC} and \ref{tab:LMC}.
Not all stars were observed in the blue, and hence there are stars for
which there are no spectral types.
These were determined by comparison of our spectra of spectral standards
to the program objects.  At our dispersion and signal-to-noise, the
presence of the TiO $\lambda 5167$ suggests that the star is K5 or later,
and the classification was based upon the strength of the TiO bands
at $\lambda \lambda$ 4761, 4954, 5167, 5448, and 5847.  If there was
no TiO present, then the relative strength of Ca~I $\lambda 4226$
and the G band were used to determined the spectral subtype in the
range K0-K5.  Strong H$\gamma$ suggested an earlier type (G-type), which 
proved to be the case for a few of the foreground dwarfs.
The luminosity criteria are quite subtle, and we relied upon our radial
velocities to guide us in assigning ``V" for foreground dwarfs, or ``I"
for supergiants.  The presence of LMC giants in our sample is
precluded by our $V$ magnitude selection criterion.

The comparison with the published spectral types for some stars in common
with Elias, Frogel, \& Humphreys (1985) [SMC], and Humphreys (1979) [LMC]
shows generally excellent agreement.  The average difference is less
than half a spectral type for the 76 stars in common.  In only three cases
the difference is 3 spectral subclasses or more; i.e., SMC-026778, which
we call K2~I but Elias et al.\ (1985) call M0~I; SMC-054708, which we call
K0~I but Elias et al.\ (1985) call M0 Iab; and LMC-178066, which we call
K7~I but Humphreys (1979) call M2~Ia.  Given the size of the discrepancy,
we speculate that these may be spectrum variables.  (In the case of the LMC
star, the identification is not certain, as only approximate coordinates had
ever been published for the Case stars that were subsequently
observed by Humphreys 1979.)

\section{Physical Parameters and Stellar Evolution}

What do these spectral types mean in terms of physical parameters?  We have
classified the stars in the traditional way, relying upon the strengths of
the TiO bands to determine the spectral type, with stronger bands leading to
a later type.  However,
the metallicity (as judged from the oxygen abundances of H~II regions)
of the LMC is about a factor of 2 lower than 
in the solar neighborhood, while the
metallicity of the SMC is about a factor of 4 times 
lower (Russell \& Dopita 1990).
Thus RSGs of the {\it same} effective temperatures in the Milky Way, LMC,
and SMC would be classified as progressively earlier in these three galaxies as
lower metal abundance weakens the TiO band strength used to classify these
stars.  
Elias et al.\ (1985) see such an effect in their comparison of
the average (median) spectral types of RSGs in SMC (M0~I), the LMC (M1~I),
and the Milky Way (M2-3~I), but attribute the change primarily to
the effect that metallicity has on the location of the (giant branch) Hayashi track, 
only secondarily to the effects on the spectral appearance of 
stars of a given effective temperature.  However, modern
evolutionary models do not show a Hayashi track for red {\it super}giants,
and so it is worth re-examining this issue.
 
We show our own histograms for the LMC and SMC supergiants in our
sample, and compare these to the distribution of spectral types for
the Milky Way taken from Table~20 in Elias et al.\ (1985).  
The medians we find are K5-7~I for the SMC, M1~I for the LMC, and M2~I for
the Milky Way.
The median spectral type in the SMC is somewhat earlier than that found by
Elias et al.\ (1985), and probably
results either from our larger sample size, or our
better completeness for early
K-type supergiants.  We find, as do Elias et al.\ (1985), that
the distribution of spectral types is more narrow in the SMC than in the
LMC or Milky Way, although we still find RSGs as late as M3~I in the SMC---just
not in large numbers.

First, let us ask if it is reasonable that this progression in average
spectral types is due solely to the effect that metallicity has on the 
relationship between spectral type and effective temperature.
In Table~\ref{tab:teff} we compare various effective temperature scales
for Galactic RSGs.
We include here the effective
temperature scale adopted by Humphreys \& McElroy (1984), based upon a
number of sources, and the Lee (1970) calibration of effective temperature
with spectral types for M-type supergiants, based primarily on a very limited
amount of ``fundamental" data (i.e., using stars with known radii).  This
work has been extended considerably in recent years by Dyck et al.\ (1996)
and Dyck, van Belle, \& Thompson (1998),
who obtained new interferometric observations at 2.2$\mu$m, and combined
these with similar data from
the literature. They provide a scale for red {\it giants}, but
consider the supergiant data to be too sparse for a calibration.  Their
supergiant data clearly lies several hundred degrees cooler than the
giant sequence (i.e., Fig.~3 in Dyck et al.\ 1996).  
Houdashelt et al.\ (2000) recently  
compared the temperatures expected from the new MARCS models to the 
Dyck et al.\ (1996) values and
found very good agreement.  We have adjusted the Dyck et al. (1996) scale
for red {\it giants}
by $-400^\circ$K (i.e., to cooler temperatures) to produce reasonable
estimates for {\it supergiants}, consistent with the temperature differences
illustrated in their Fig.~3.  (See also discussion following Bessell 1998.)
Comparing all of these values have led to a somewhat arbitrary effective
temperature scale which we adopt here, noting that the present uncertainties
prevent a more definitive answer at this time.  What sort of change is expected on the basis of metallicity?
Improved stellar atmospheres applicable to RSGs are
under construction (Gustafsson et al.~2003, Plez 2003) but until these
are generally available we turn to the ``Atlas 9" model atmospheres
of Kurucz (1992) to help answer this question\footnote{We note that these
Kurucz (1992) models are the primary component of the compilation of 
``standard" synthetic spectra available on the Web by 
T. Lejeune and collaborators, particularly in the realm of RSGs; 
see Fig.~1 of Lejeune, Cusinier, \& Buser (1998). Although Bessell 
et al.~ (1989, 1991) have published a few models appropriate to RSGs
at Galactic and SMC-like metallicities, they lack LMC-like metallicities
and the grid points are sparse, causing us to adopt the Kurucz (1992) models,
despite their less exact treatment of molecules.}.
Although Oestreicher \& Schmidt-Kaler (1998) find that the Atlas 9
models significantly underestimate the amount of molecular absorption for
{\it some} lines in late-type stars, we show that there is pretty good
agreement in Fig.~\ref{fig:kurucz}, where we compare the coolest of Kurucz
(1992) models to the spectra of three of our spectral standards. The Kurucz
(1992) models correspond to solar metallicity and $\log g=0.0$, which
is appropriate for a massive supergiant.\footnote{We expect 
that $\log g$ will vary
from about $-0.3$ 
(20$\cal M_\odot$, $M_{\rm bol}=-8.0$, $\log T_{\rm eff}=3.50$) to $-0.6$
(40$\cal M_\odot$, $M_{\rm bol}=-9.5$, $\log T_{\rm eff}=3.55$).
Thus the $\log g=0.0$ Kurucz (1992)
models are the most appropriate ones available
for RSGs. Fortunately, the strengths of the TiO bands in general are not
sensitive to the exact choice of $\log g$; see Schiavon \& Barbuy (1999).}
We see that the 3500$^\circ$K
model shows TiO lines that are roughly comparable with what is seen
in M0-M2~I supergiants, consistent with the effective temperature scale
we adopted in Table~\ref{tab:teff}.  Similarly, the TiO bands in the
4000$^\circ$K model are similar to that of the K2.5~I standard, also in accord
with the effective temperature scale adopted above.
The spectra are
plotted in log units in order to facilitate comparison of band depths
without the subjective task of normalization.   The continuum fluxes
of the stellar spectra have been adjusted by comparison with stars of
similar spectral types from the Jacoby, Hunter, \& Christian (1984) atlas,
and so the relative fluxes are only approximate; what matters is the line
depths.

We next investigate the effect the Kurucz (1992) models predict for a
change in metallicity from Galactic to that of the SMC, where we have
observed the spectral types change from M2~I to K5~I.  The red curve in
Fig.~\ref{fig:kurucz} shows the Kurucz (1992) model for a 3500$^\circ$K
supergiant computed with an abundance $\log Z/Z_o=-0.5$, while the blue
curve corresponds to $\log Z/Z_o=-1.0$.  The metallicity of SMC should be
intermediate between these two values.  We see that metallicity alone has
changed the line depths to be intermediate between the 3750$^\circ$K and
4000$^\circ$ models.  Thus the change in metallicity from the Milky Way
to that of the SMC is likely to weaken the appearance of the TiO spectral
lines by an amount corresponding to $+250^\circ$K to $+500^\circ$K.  
This is consistent
with the $\sim +300^\circ$K temperature difference between (Galactic) M2~I and
K5-7~I stars. Thus the effective temperature scale at lower metallicity
will be cooler; i.e., an SMC M0~I star would be 300 cooler than a Galactic
M0~I star. A  more quantitative comparison requires 
higher resolution
synthetic spectra and a finer temperature and metallicity grid, and these
will soon be available for such tests
from the MARCS group (Gustafsson et al.\ 2003, 
and Plez 2003)\footnote{Bernard Plez (private communication) kindly gave us
a chance to examine some of his models. Unfortunately, the surface gravities
were $\sim 10\times$ that expected for a supergiant, so the application to
the stars we discuss here is not straight-forward.  We will note that his
spectra apply a considerably warmer temperature scale (and considerably
more compressed) than what we adopt here.
At first blush, the warmer scale appears to be in 
disagreement with the fundamental data of
Lee (1970) and Dyck et al.~(1996).  The models do not show much of an
effect with metallicity, but a more careful comparison done with 
absolute spectrophotometry, with more appropriate
surface gravities, is needed.}. In the meanwhile, we will adopt an effective temperature
scale for the Magellanic Cloud RSGs that is 300$^\circ$ cooler for the SMC, and 100$^\circ$ cooler for the LMC, compared with the Milky Way, consistent with
the average change in spectral type we observe.

We can provide an additional check on this by examining the intrinsic colors.
It is generally recognized that 
$(V-R)_o$ is a good effective temperature indicator for cool stars, while
$(B-V)_o$ is sensitive both to effective temperature and to surface
gravity (e.g., Lee 1970, Massey 1998b, Oestreicher \& Schmidt-Kaler
1999).  
In Table~\ref{tab:kurucz} we give the expected $(B-V)_o$ and
$(V-R)_o$ colors
as a function of effective temperature and metallicity computed from
the Kurucz (1992) Atlas 9 models, where we have adopted the description of the
{\it B} and {\it V} band-passes from Buser \& Kurucz (1992),
and that of the Kron-Cousins {\it R} bandband from Bessel (1983).  We include
in Table~\ref{tab:kurucz} the approximate corresponding spectral types for
Galactic stars, using Table~\ref{tab:teff}.
We see that there is very little change in color with metallicity
for the ``warmer" models (3750$^\circ$K to 4000$^\circ$K).  For the 
3500$^\circ$K model there is no change from the Milky Way to the LMC, but
that we expect that $(V-R)_o$ will be significant larger (0.07~mag) in
the SMC.  

How do these colors compare to the observed photometry?  In Table~\ref{tab:colors} we give the
average $(B-V)_o$ and $(V-R)_o$ colors for our spectral types, where
we have corrected the observed colors in Tables~\ref{tab:SMC} and
\ref{tab:LMC} by the average reddenings as indicated in the footnote
to the table.  We have used the arithmetic means at each spectral type,
after rejecting the highest and lowest values in producing these averages.
We do not list colors for any spectral types with 3 or fewer representatives.

For the LMC there is relatively good agreement:
we expect an LMC M0~I star to have $T_{\rm eff}=3500^\circ$K (i.e.,
100$^\circ$ cooler than the value listed in Table~\ref{tab:teff})
and the Kurucz (1992) model atmospheres predict a $(V-R)_o$ color
of 0.92.  We observe $0.94\pm0.01$ (Table~\ref{tab:colors}).  However, for
the SMC the agreement is poor between the Kurucz (1992) $(V-R)_o$ colors
and those observed. If we correct the Galactic scale
by $-300^\circ$K as argued above, a 3500$^\circ$K SMC star should 
have a spectral
type of K5~I. 
The models then predict a $(V-R)_o$ color of 0.99.  But, what we actually
observe is a $(V-R)_o$ color of 0.84.  If we had made {\it no} correction
to the Galactic effective temperature scale then the broad-band
colors would be in pretty good agreement.  Have we fooled ourselves in
making this correction?  Possibly.  However, Bessell et al.~(1989) has
published a few models applicable to cool supergiants. We give their
$(V-R)_o$ colors in Table~\ref{tab:bessell}.  Their SMC-like
metallicity ($Z=-0.5$) supergiant model predict a $(V-R)_o$ color of 0.84
at 3500$^\circ$K ($\log$g=-0.26), 
in excellent agreement with the observed colors if we apply our temperature
correction. Similarly, their $T_{\rm eff}=3350^\circ$K model predicts
a $(V-R)_o$ color of 0.92.  Applying our correction, we would
expect this temperature would correspond to an SMC star of spectral type
K7-M0~I, and indeed we find an observed color of 0.88-0.94, in good agreement.
LMC-like metallicity models were not computed by Bessell et al.~(1989),
limiting the degree we can make this comparison, but we note that their
Galactic $(V-R)_o$ colors are significantly bluer than the Kurucz (1992)
models would predict (e.g., 0.74 vs. 0.92 at $T_{\rm eff}=3350^\circ$K).
A finer grid of 
higher-resolution synthetic spectra appropriate to cool supergiants is needed
before a metallicity-dependent effective temperature scale can be reliably
derived.

Let us next compare the distribution of stars in the H-R diagram to that
predicted by stellar evolutionary models. We use the ``corrected" temperatures
for the spectral types, as defined above.
For stars without spectral types, we can use the $(V-R)_o$ to determine
an effective temperature.  Comparison of our measured colors for the
stars with spectral types produces two linear relations:
$$\log T_{\rm eff}=3.899-0.4085\times (V-R)_o {\rm (SMC)}$$
$$\log T_{\rm eff}=3.869-0.3360\times (V-R)_o {\rm (LMC)}.$$

The conversion from the
adopted effective temperatures to bolometric corrections is made by 
using the relation of Slesnick, Hillenbrand, \& Massey (2002), which is 
primarily a fit to the bolometric corrections as a function of effective
temperatures tabulated by Humphreys \& McElroy (1984).   We have included
the adopted $T_{\rm eff}$ and $M_{\rm bol}$ in Tables~\ref{tab:SMC} and
\ref{tab:LMC}.  

As the Massey (2002a) photometric survey 
was limited in area and could conceivably
suffer from saturation for the most luminous supergiants, we should also
consider other stars that have been spectroscopically confirmed as Magellanic
Cloud RSGs.  We list these in Tables~\ref{tab:smcother} and \ref{tab:lmcother}.
In the case of the
SMC we have excellent cross-reference to the spectral types of 
Elias et al.\ (1985) thanks to the good coordinates provided by Sanduleak
(1989).  However, cross-referencing
to the spectral types of Humphreys (1979) was
more of a challenge as only approximate coordinates were provided in the
Case objective prism survey (Sanduleak \& Philip 1977)
from which Humphreys (1979) drew her sample
for spectroscopy.  Thus Massey (2002a) gives all cross-identifications for the
LMC stars as tentative, and there were a number of stars for which several
possible matches were a possibility.  Thus there may be other previously
observed LMC RSGs that we have incorrectly adopted as identical to our
stars in Table~\ref{tab:LMC}.

We show the H-R diagrams in Figs.~\ref{fig:smchrds} and \ref{fig:lmchrds},
where we include the evolutionary tracks both of the Geneva and Padova
groups (Schaerer et al.\ 1993; Meynet et al.\ 1994; Fagotto et al.\ 1994).
We see here a very interesting effect, namely that {\it none} of these
evolutionary tracks produce RSGs that are as cool and as luminous 
as observed
in the Magellanic Clouds.  Although the agreement is good at 12$\cal M_\odot$
masses, at higher masses the tracks simply do not go far enough to the
right (cool temperatures) to produce the RSGs that we observe.
It would appear that the RSG sequence extends up
to perhaps 40$\cal M_\odot$, but that those tracks simply do not go 
sufficiently far to the right in the diagram.  Massey (2002b, 2003) finds
that the identical problem exists for Galactic RSGs even when adopting
the effective temperature scale and luminosities of Humphreys (1978).

This discrepancy has also been suggested by the poor match of
synthetic ``starburst" spectra to observations of the integrated light
of various stellar populations (e.g., Mayya 1997, Oliva \& Origlia 1998, Origlia et al.\ 1999).  It is quite apparent even if one looks only at the 
broad-band colors derived by Lejeune \& Schaerer (2001).  For instance,
if one considers the 40$\cal M_\odot$ evolutionary track computed with
$Z=0.004$ and a normal mass-loss rate by Charbonnel et al.\ (1993), which
is shown in Fig.~4(a), Lejeune \& Schaerer 2001) compute $B-V=0.244$ and
$V-R=0.157$ at the coolest extension of the track. These colors correspond
to a mid-F type supergiant. 

What can account for the problem with the evolutionary tracks?  
One possibility is the difference
that the treatment of convection can make in the evolutionary tracks.
This is 
illustrated in Fig.~9 of Maeder \& Meynet (1987), where they compare the
older B\"{o}hm-Vitense (1981) mixing length (1.5 times the local pressure
scale height) with a more accurate treatment that includes the
effects of turbulent pressure and acoustic flux and has the mixing length
proportional to the density scale height. 
Although the physics is better,
the result is that the evolutionary tracks no longer produce
RSGs that are as luminous and cool as earlier models had.  However, the
Padova models reply upon the older B\"{o}hm-Vitense (1981) prescription,
albeit it with a mixing length of 1.63 times 
the pressure scale height, and these
too suffer from the same problem, as shown in Figs.~\ref{fig:smchrds} and \ref{fig:lmchrds}.  Maeder \& Meynet (1987) were certainly aware of the
mis-match between theory and observation, and expressed the hope that 
``complete stellar models" (i.e., ones which included the 
extended atmospheres caused by stellar winds) would some day alleviate
the problem.  Such winds would make the star larger than the (purely)
interior models would suggest, lowering the effective temperature.
In the meanwhile, this discrepancy has been generally
ignored by the users of these models.

What if we had ignored the effective temperature corrections?  In 
Fig.~\ref{fig:oldhrds} we compare the H-R diagrams for the SMC and LMC
adopting the Galactic spectral type to effective temperature calibration from
Table~\ref{tab:teff},  and computing the corresponding color to effective
temperature equation. 
It is clear that if we
made no correction that the RSGs would be of considerably higher luminosity
in the LMC than in the SMC.  In the left-hand side of 
Fig.~\ref{fig:mbolhists} we compare the 
distribution of bolometric luminosities for RSGs both with and without
these corrections.  We see that without the corrections the number of RSGs
in the SMC drops very abruptly with increasing luminosity compared to the
LMC. 
This runs
counter to the expected evolutionary effect that at lower metallicities higher
mass (luminosity)
stars should spend a greater fraction of their He-burning lifetimes as
RSGs rather than WRs since mass-loss rates will be lower at low metallicities
(see Maeder et al.\ 1980 and Maeder \& Conti 1994).
Indeed, Massey (1998b) found a smooth decrease in the numbers of the highest
luminosity RSGs as metallicity increased from NGC~6822 (SMC-like) to M~33
(LMC-like) to M~31 (higher than solar).  When we make the correction for
effective temperature, however, the luminosity functions become very similar (right-hand side of Fig.~\ref{fig:mbolhists}.
Thus either a significant correction to
the Galactic $T_{\rm eff}$ scale is needed for SMC RSGs, as
we have made above, or else there is an unexpected
absence of higher luminosity RSGs in the SMC.

We note that in the case of the H-R diagrams with the corrected
temperatures that we expect that
the most luminous RSGs come from stars with initial masses of about
40$\cal M_\odot$.  This is consistent with so-called upper luminosity
limit described by Humphreys \& Davidson (1979), and explained by
Lamers (1997): the ``modified" Eddington limit should prevent stars
with luminosities above $M_{\rm bol}\sim -10$ from evolving to the right
in the H-R diagram. This limit should, if anything, be slightly higher
at lower metallicities, since the opacities will be lower, and thus is
consistent with our corrected temperatures.

Perhaps one of the most interesting things to be apparent in the H-R diagrams
is that there is a very smooth decrease in effective temperature with
increasing luminosity, whether or not the temperature corrections are applied.
{\it The higher luminosity RSGs are invariably of cooler effective temperatures.}  This tight sequence is obviously not reproduced by the
stellar evolutionary models.  Explaining this simple sequence provides an
important challenge to stellar evolutionary theory.

Finally, let us briefly reconsider the ratio of blue-to-red supergiants (B/R)
in the SMC and the LMC.  As emphasized in Massey (2002), one needs to be
careful in what one is counting for this ratio to have much meaning.
We need to include K-type as well as M-type stars, but would like to
exclude stars earlier than K-type as the degree of foreground contamination
increases at intermediate colors.  We adopt the same convention as Massey (2002a), namely $(V-R)_o>0.6$, corresponding to a star of $\log T_{\rm eff}=3.66$
(4600$^\circ$K, or late G-type).  We also restrict ourselves to counting only
stars with $M_{\rm bol}<-7.5$, as less luminous than this there is a chance
of contamination by intermediate-mass asymptotic giant branch stars
(Brunish, Gallagher, \& Truran 1986).  In counting stars we include all
of the sufficiently luminous RSGs in Tables~\ref{tab:SMC} and \ref{tab:LMC},
plus a fraction of the other red stars from Massey (2002a). Our spectroscopy
suggests that 11\% of the red stars seen towards the
SMC, and 5.3\% of the red stars seen towards the LMC
are foreground, so we count only 89\% and 94.7\% of the remainder\footnote{Since we have adopted a new conversion between $(V-R)_o$ and $\log T_{\rm eff}$, we
started with the complete photometric catalog (Table 3) of Massey (2002a)
rather than the list of just the red, luminous stars (Table 9), but
the differences are small.}.  We find that we expect about 90 RSGs in the
SMC sample, and 234 RSGs in the LMC sample.
For the blue stars, we use the numbers given in Table~10 of Massey (2002),
i.e., all of the stars in the SMC and LMC areas surveyed that meet the
criteria $M_{\rm bol}<-7.5$ and $(B-V)_o<0.14$, where the latter 
roughly corresponds to the color of an A9~I star ($\log T_{\rm eff}=3.9$).  
We then count
1484 blue supergiants in the SMC, and 3164 blue supergiants in the LMC, 
although these numbers are considerably uncertain given the difficulty in
converting photometry to $\log T_{\rm eff}$ and $M_{\rm bol}$ for hot
stars. (See, for example, Massey et al.\ 1998a.) 
The derived B/R values are thus 16 for the
SMC, and 14 for the LMC, essentially identical.   Massey (2002) notes that
a large (factor of 3) difference is found if only M-type stars are counted.
Thus the fact that the stellar models fail to reproduce the ``observed" B/R
value (Langer \& Maeder 1995) may be in large part due to the differences
in how the ``observed" ratios have actually been counted.  The slightly
different approach here has changed the B/R ratio given by Massey (2002a)
by nearby a factor of 2 in itself, and thus we again emphasize the large
``observational" uncertainty in such a census, as the derived ratio is 
highly sensitive to the conversions to bolometric luminosity.

We can compare this number to the predicted from stellar models.
For this comparison we follow the advice offered by Schaerer \& Vacca
(1998) to determine the number of stars from the model by integrating
the initial mass function over closely-spaced isochrones rather than
by
integrating over the coarsely-spaced mass tracks.  
The SMC-like $Z=0.004$
Geneva
models with normal mass-loss rates (Charbonnel et al.\ 1993) predict
a blue-to-red supergiant ratio of 54, while the enhanced mass-loss
models (Meynet et al.\ 1994)
would expect a blue-to-red supergiant ratio of 36.  The LMC-like
$Z=0.008$ Geneva models with normal mass-loss rates (Schaerer et al.\ 1993)
predict a B/R value of 10, while those using enhanced mass-loss rates 
(Meynet et al.\ 1994)
predict a B/R value of 3.  Thus, the lower metallicity models (SMC-like)
predict
a much higher ratio of B/R than what is observed, 
while the intermediate
metallicity models (LMC-like)  predict a somewhat lower value. 
Given that we have earlier shown
that the models fail to reproduce the location of RSGs in the H-R 
diagram, the disagreement is not surprising.
Maeder \& Meynet (2001) find that more RSGs are produced in the models
at SMC-like metallicities when rotation is included.
Comparisons with the new rotation
models that cover a range of metallicities
will be of great interest.

\section{Summary and Conclusions}

We have examined samples of red stars seen towards the SMC and LMC.  Our
spectroscopy has been able to determine membership based upon radial
velocity information; we find that the contamination by foreground red
dwarfs is about 11\% in the SMC sample and 5.3\% in the LMC sample.

Classification of our spectra confirms that there is a progression in the
average spectral type of RSGs with metallicity.  RSGs in the
SMC (which is the lowest in metallicity) have an average spectral type
of K5-7~I.  Nevertheless, there are a few SMC RSGs as late as M3 in our sample.
In the LMC have an average type of M1~I, while those in the Milky
Way have an average type of M2~I.  
At lower metallicity the appearance
of the TiO lines (used as the primary classification criterion) should be
weaker, and examination of the Kurucz (1992) Atlas 9 models suggest
that this effect is probably sufficient in itself to account for the shift
in average spectral types observed.  If so, then
RSGs in the SMC are about 300$^\circ$ cooler than Galactic
stars of the same spectral types,  while RSGs in the LMC are
about 100$^\circ$ cooler.  
The $(V-R)_o$ colors predicted by the Kurucz (1992)
models do not agree with this
conclusion, but other models (e.g., Bessell 1989) show better agreement
with the SMC data although lack the LMC-like metallicity we would
need to draw conclusions. Good resolution ($<10$\AA) synthetic spectra for
red supergiants ($\log g=0.0$) covering a range of metallicities is needed
(along with good spectrophotometry)
to address this discrepancy.

We find that none of the stellar evolutionary models produce RSGs that are
as red and luminous as observed in the Magellanic Clouds. This discrepancy
may be due to the treatment of convection in the evolutionary models, or could
simply be due the lack of inclusion of that stellar winds have in increasing
the atmosphere extent (leading to a decrease in the effective temperature)
in the stellar models.  Nevertheless the location of RSGs compared to the
evolutionary tracks suggest that the most luminous RSGs have evolved from
stars with initial masses of 40$\cal M_\odot$, in accord with previous
studies.  We show that ignoring the temperature correction described above
would lead to an under-abundance of high luminosity SMC RSGs.

There is a very tight sequence in the H-R diagram in which the higher
luminosity RSGs are of lower effective temperatures.  Matching this sequence
will be an important test of future stellar models.

The blue-to-red supergiant ratio does not appear to be significantly different
in the SMC and in the LMC, although there is still considerable 
uncertainty in the number of blue supergiants in our sample.  This work has
underscored the point made by Massey (2002a) that the
B/R value is very dependent upon how stars are counted, and thus disagreements
with the predictions of stellar evolutionary models have to be carefully
evaluated.  Using the non-rotation Geneva models, we find that the
SMC-like ($Z=0.004$) models 
predict too large a value for B/R, while the LMC-like
($Z=0.008$) models 
predict too small a value.  Given the fact that the models fail
to produce high luminosity red supergiants, such disagreements are not
surprising.  The effects that rotation will have on the predicted B/R
ratio as a function of metallicity remain unclear.  As Maeder \& Meynet
(2000) describe, the additional
mixing caused by rotational instabilities would tend to produce few RSGs,
while on the other hand higher rotation will lead to increase mass loss, and 
this would tend to produce more RSGs. The result is that it is still unclear
what affect, if any, the complete inclusion of rotation will have on the
predictions of B/R ratios as a function of 
metallicities, although  Maeder \& Meynet (2001) 
find that at a SMC-like metallicity including rotation will
lower the predicted B/R ratio, which goes in the correction direction.
Eggenberger,  Meynet \& Maeder (2002) compare
the observed B/R ratios of clusters to those of models, but as discussed
extensively by Massey (2002, 2003) the ``B/R" ratio in a quasi-coeval
situation will be quite different than in a mixed-age population, such
as what we consider here.

\acknowledgments

This work was supported by NSF Grant AST0093060.  We are grateful, as always,
to the kind hospitality and excellent support by the technical staff at
CTIO.  We also benefited from correspondence with Ken Hinkle and
Daniel Schaerer.
Deidre Hunter offered useful comments on an early draft of this manuscript.

\clearpage

\begin{deluxetable}{l c c c c c c r r l l l}
\tabletypesize{\footnotesize}
\tablewidth{0pc}
\tablenum{1}
\tablecolumns{12}
\tablecaption{\label{tab:SMC}Red Stars Seen Towards the SMC\tablenotemark{a}}
\tablehead{
\multicolumn{10}{c}{}
&\multicolumn{2}{c}{Spectral Type} \\ \cline{11-12} 
\colhead{Star}
& \colhead{$\alpha_{2000}$}
& \colhead{$\delta_{2000}$}
& \colhead{V}
& \colhead{B-V}
& \colhead {V-R}
& \colhead {log $T_{\rm eff}$\tablenotemark{b}}
& \colhead {$M_{\rm bol}$\tablenotemark{b}}
& \colhead {RV\tablenotemark{c}}
& \colhead {Member?}
& \colhead {New}
& \colhead {Lit.\tablenotemark{d}}
}
\startdata
008324&00 47 16.84&$-$73 08 08.4&13.08& 1.64& 0.85& 3.565& -7.32& 134.4&SMC &K0: I   & \nodata      \\
008367&00 47 18.11&$-$73 10 39.3&12.46& 1.40& 0.93& 3.531& -8.58& 127.9&SMC &K7 I    &  \nodata    \\
008930&00 47 36.94&$-$73 04 44.3&12.68& 2.00& 1.06& 3.531& -8.36& 131.6&SMC &K7 I    &M1 Ia  \\
009766&00 48 01.22&$-$73 23 37.5&12.95& 1.29& 0.86& 3.531& -8.09& 141.6&SMC &K7 I    & \nodata      \\
010889&00 48 27.02&$-$73 12 12.3&12.20& 2.00& 1.06& 3.531& -8.84& 138.4&SMC &K7 I    &M0 Ia  \\
011101&00 48 31.92&$-$73 07 44.4&13.54& 1.69& 0.99& 3.531& -7.50& 146.4&SMC &K7 I    & \nodata      \\
011709&00 48 46.32&$-$73 28 20.7&12.43& 1.79& 0.94& 3.531& -8.61& 140.4&SMC &K7 I    &K5-M0I \\
011939&00 48 51.83&$-$73 22 39.3&12.82& 1.81& 1.00& 3.518& -8.59& 131.8&SMC &M0 I    &  \nodata     \\
012322&00 49 00.32&$-$72 59 35.7&12.44& 1.93& 1.03& 3.531& -8.60& 149.0&SMC &K7 I    &M0 Ia  \\
012572&00 49 05.25&$-$73 31 07.8&11.66& 1.45& 0.76& 3.602& -8.35& 228.5&SMC & \nodata&  \nodata     \\
012707&00 49 08.23&$-$73 14 15.5&13.40& 1.77& 1.00& 3.503& -8.52& 162.6&SMC & \nodata&  \nodata       \\
013472&00 49 24.53&$-$73 18 13.5&11.73& 1.77& 0.85& 3.531& -9.31& 137.6&SMC &K7: I   &K0-K5I \\
013740&00 49 30.34&$-$73 26 49.9&13.47& 1.77& 0.96& 3.531& -7.57& 156.4&SMC &K7 I    &  \nodata      \\
013951&00 49 34.42&$-$73 14 09.9&13.00& 1.79& 0.93& 3.531& -8.04& 125.0&SMC &K7 I   &  \nodata      \\
015510&00 50 06.42&$-$73 28 11.1&12.59& 1.90& 0.95& 3.518& -8.82& 163.0&SMC &M0 I   &M0 I   \\
017656&00 50 47.22&$-$72 42 57.2&12.66& 1.69& 0.90& 3.568& -7.70& 134.0&SMC &K0-5 I & \nodata       \\
018592&00 51 03.90&$-$72 43 17.4&11.39& 1.82& 0.95& 3.568& -8.97& 152.3&SMC &K0-2 I &K5-M0I \\
019551&00 51 20.23&$-$72 49 22.1&12.98& 1.04& 0.83& 3.568& -7.38& 145.4&SMC &K2 I   & \nodata       \\
019743&00 51 23.28&$-$72 38 43.8&13.45& 1.67& 1.05& 3.544& -7.29& 138.2&SMC &K5 I   &M0 Iab \\
020133&00 51 29.68&$-$73 10 44.3&12.33& 1.95& 1.03& 3.518& -9.08& 170.4&SMC &M0 I   &M0 Iab \\
020612&00 51 37.57&$-$72 25 59.5&12.97& 1.64& 0.82& 3.544& -7.77& 154.9&SMC &K5 I   &K5-M0  \\
023463&00 52 26.51&$-$72 45 15.6&12.44& 1.35& 0.90& 3.568& -7.92& 157.9&SMC &K0-5 I & \nodata       \\
023700&00 52 30.69&$-$72 26 46.8&13.09& 1.67& 0.85& 3.568& -7.27& 149.8&SMC &K0-2 I & \nodata       \\
025550&00 53 02.85&$-$73 07 45.9&13.35& 1.67& 0.94& 3.568& -7.01& 136.8&SMC &K2 I   & \nodata       \\
025879&00 53 08.87&$-$72 29 38.6&11.91& 1.77& 0.88& 3.531& -9.13& 134.5&SMC &K7 I   &M0 Ia  \\
025888&00 53 09.04&$-$73 04 03.6&12.08& 1.82& 0.95& 3.538& -8.80& 159.1&SMC &K5-7 I &M0 Ia- \\
026402&00 53 17.81&$-$72 46 06.9&12.78& 1.05& 0.75& 3.568& -7.58& 148.4&SMC &K0-2 I & \nodata       \\
026778&00 53 24.56&$-$73 18 31.6&12.78& 1.55& 0.95& 3.568& -7.58& 153.0&SMC &K2 I   &M0 Iab \\
027443&00 53 36.44&$-$73 01 34.8&12.75& 1.86& 1.01& 3.531& -8.29& 140.3&SMC &K7 I   & \nodata       \\
027945&00 53 45.74&$-$72 53 38.5&12.94& 1.57& 0.80& 3.552& -7.61& 135.4&SMC &K3-5 I & \nodata       \\
030135&00 54 26.90&$-$72 52 59.4&12.84& 1.68& 0.78& 3.568& -7.52& 150.8&SMC &K0-2 I & \nodata       \\
030616&00 54 35.90&$-$72 34 14.3&12.22& 1.85& 0.92& 3.531& -8.82& 140.4&SMC &K7 I   &M0 Iab \\
032188&00 55 03.71&$-$73 00 36.6&12.40& 1.75& 0.86& 3.544& -8.34& 154.1&SMC &K5 I   & \nodata       \\
033610&00 55 26.82&$-$72 35 56.2&12.60& 1.75& 0.91& 3.531& -8.44& 157.4&SMC &K7 I   &M0 Iab \\
034158&00 55 36.58&$-$72 36 23.6&12.79& 1.78& 0.95& 3.531& -8.25& 139.0&SMC &K7 I   &K5-M0  \\
035231&00 55 55.10&$-$72 40 30.4&12.02& 1.32& 0.66& 3.568& -8.34& 151.8&SMC &K2 I   & \nodata       \\
037994&00 56 43.55&$-$72 30 15.0&12.65& 1.68& 0.97& 3.531& -8.39& 148.6&SMC &K7 I   &K5-M0  \\
041778&00 57 56.45&$-$72 17 33.3&12.52& 1.08& 0.82& 3.531& -8.52& 178.9&SMC &K7 I   & \nodata       \\
042319&00 58 06.61&$-$72 20 59.8&13.09& 1.90& 0.94& 3.556& -7.41& 184.9&SMC &K2-5 I & \nodata       \\
042438&00 58 08.71&$-$72 19 26.7&13.20& 1.59& 0.87& 3.552& -7.35& 176.5&SMC &K3-5 I &\nodata  \\
043219&00 58 23.30&$-$72 48 40.7&13.06& 1.84& 0.94& 3.518& -8.35& 135.7&SMC &M0 I   &M0 Iab \\
043725&00 58 33.21&$-$72 19 15.6&13.50& 1.56& 0.96& 3.544& -7.24& 182.7&SMC &K5 I   &\nodata  \\
044719&00 58 53.33&$-$72 08 35.3&12.98& 1.53& 0.82&  \nodata &  \nodata &  95.4&Fgd?&K5 V?  &  \nodata      \\
044724&00 58 53.54&$-$72 40 38.7&11.78& 1.59& 0.87&  \nodata &  \nodata &  55.5&Fgd &Dwarf  & \nodata       \\
044763&00 58 54.44&$-$72 41 40.8&12.73& 1.28& 0.82&  \nodata &  \nodata &  18.2&Fgd &Dwarf  & \nodata       \\
045378&00 59 07.16&$-$72 13 08.6&12.93& 1.56& 0.92& 3.544& -7.81& 179.9&SMC &K5 I   &K5 I   \\
045850&00 59 16.90&$-$72 25 10.9&12.88& 1.76& 0.87& 3.568& -7.48& 141.8&SMC &K0-5 I &K5-M0  \\
046497&00 59 31.33&$-$72 15 46.4&12.40& 1.98& 0.99& 3.505& -9.46& 166.3&SMC &M1 I   &M0 Ia- \\
046662&00 59 35.04&$-$72 04 06.2&12.90& 1.88& 1.07& 3.491& -9.54& 180.2&SMC &M2 I   &M0 Ia  \\
046910&00 59 40.58&$-$72 20 55.9&12.82& 1.75& 0.85& 3.552& -7.73& 160.2&SMC &K3-5 I &M0 Ia  \\
047757&01 00 00.63&$-$72 19 40.2&12.52& 1.87& 1.02& 3.505& -9.34& 161.1&SMC &M1 I   &K5-M0  \\
048122&01 00 09.42&$-$72 08 44.5&12.19& 1.78& 0.89& 3.556& -8.31& 172.8&SMC &K3 I   & \nodata       \\
049033&01 00 30.43&$-$71 58 24.7&12.50& 1.82& 0.91& 3.544& -8.24& 160.2&SMC &K5 I   &M0 I   \\
049428&01 00 40.32&$-$72 35 58.8&12.97& 1.73& 0.87& 3.544& -7.77& 134.4&SMC &K0-7 I &K5 I   \\
049478&01 00 41.56&$-$72 10 37.0&12.17& 1.81& 0.99& 3.518& -9.24& 177.1&SMC &M0 I   &K5 Ia  \\
049990&01 00 54.13&$-$72 51 36.3&12.20& 1.66& 0.85& 3.544& -8.54& 186.8&SMC &K5 I   &K5 Ia  \\
050237&01 01 00.31&$-$72 13 41.6&12.91& 1.62& 0.84& 3.556& -7.59& 179.2&SMC &K2-5 I &K5 I   \\
050348&01 01 03.26&$-$72 04 39.4&12.92& 1.44& 0.84& 3.531& -8.12& 179.3&SMC &K7 I   & \nodata       \\
050360&01 01 03.58&$-$72 02 58.5&13.09& 1.61& 0.86& 3.544& -7.65& 163.7&SMC &K5 I   & \nodata       \\
050840&01 01 15.99&$-$72 13 10.0&12.57& 1.95& 1.02& 3.499& -9.55& 179.9&SMC &M1-2 I & \nodata       \\
051000&01 01 19.92&$-$72 05 13.1&12.89& 1.66& 0.85& 3.544& -7.85& 177.7&SMC &K5 I   &K5 I   \\
051265&01 01 26.89&$-$72 01 41.3&12.87& 1.51& 0.86& 3.552& -7.68& 159.1&SMC &K3-5 I & \nodata       \\
051694&01 01 37.77&$-$71 54 16.3&11.83& 1.19& 0.72&  \nodata &  \nodata &  17.1&Fgd &G V    & \nodata       \\
051906&01 01 43.57&$-$72 38 25.1&13.02& 1.29& 0.83& 3.544& -7.72& 148.1&SMC &K5 I   & \nodata       \\
052334&01 01 54.16&$-$71 52 18.8&12.89& 1.94& 0.99& 3.531& -8.15& 165.5&SMC &K7 I   &M0 Iab \\
052389&01 01 55.43&$-$72 00 29.5&12.85& 1.60& 0.91& 3.531& -8.19& 183.8&SMC &K7 I   &K2 I   \\
053557&01 02 23.71&$-$72 55 21.2&12.72& 1.77& 0.91& 3.531& -8.32& 170.8&SMC &K7 I   &M0 I   \\
053638&01 02 25.83&$-$72 38 56.9&13.16& 1.83& 0.89& 3.544& -7.58& 153.4&SMC &K2-7 I &  \nodata      \\
054111&01 02 37.22&$-$72 16 25.1&12.55& 1.74& 0.87& 3.568& -7.81& 153.6&SMC &K0-5 I &K5-M0  \\
054300&01 02 42.12&$-$72 37 29.1&13.02& 1.74& 0.89& 3.568& -7.34& 153.8&SMC &K0-5 I & \nodata       \\
054414&01 02 44.82&$-$72 01 51.9&12.93& 1.65& 0.85& 3.552& -7.62& 174.0&SMC &K3-5 I & \nodata       \\
054708&01 02 51.37&$-$72 24 15.5&12.82& 1.81& 0.91& 3.540& -8.01& 136.8&SMC &K0 I   &M0 Iab \\
055188&01 03 02.38&$-$72 01 52.9&14.96& 2.25& 1.48& 3.491& -7.48& 176.8&SMC &M2 I   &  \nodata      \\
055275&01 03 04.34&$-$72 34 12.8&12.91& 1.70& 1.02& 3.525& -8.31& 212.2&SMC &K7-M0 I&K5-M0  \\
055355&01 03 06.43&$-$72 28 35.1&12.45& 1.86& 0.95& 3.525& -8.77& 137.6&SMC &K7-M0 I&K5-M0  \\
055462&01 03 08.80&$-$72 44 55.1&12.21& 1.38& 0.83&  \nodata &  \nodata &   3.2&Fgd &Dwarf  & \nodata       \\
055470&01 03 08.88&$-$71 55 50.8&13.12& 1.75& 0.86& 3.552& -7.43& 145.1&SMC &K3-5 I & \nodata       \\
055560&01 03 10.93&$-$72 18 32.9&12.96& 1.66& 0.90& 3.552& -7.59& 159.2&SMC &K3-5 I &K5-M0  \\
055681&01 03 12.98&$-$72 09 26.5&12.52& 1.65& 0.96& 3.478&-10.53& 182.3&SMC &M3 I   &M0-M1  \\
055933&01 03 18.56&$-$72 06 46.2&12.53& 0.98& 0.75& 3.552& -8.02& 178.8&SMC &K3-5 I &  \nodata      \\
056389&01 03 27.61&$-$72 52 09.4&11.85& 2.01& 1.01& 3.538& -9.03& 157.0&SMC &K5-7 I &M0 I   \\
056732&01 03 34.30&$-$72 06 05.8&12.86& 1.53& 0.94& 3.531& -8.18& 183.1&SMC &K7 I   & \nodata       \\
057386&01 03 47.35&$-$72 01 16.0&12.71& 1.57& 0.85& 3.552& -7.84& 170.3&SMC &K3-5 I &K5-M0  \\
057472&01 03 48.89&$-$72 02 12.7&12.80& 1.83& 0.88& 3.538& -8.08& 175.9&SMC &K5-7 I &K5-M0  \\
058100&01 04 01.64&$-$72 08 25.2&11.14& 1.42& 0.76&  \nodata &  \nodata &  -0.3&Fgd &K2-7 V & \nodata       \\
058149&01 04 02.77&$-$72 05 27.7&12.96& 1.48& 0.85& 3.556& -7.54& 177.1&SMC &K2-5 I &K5-M0  \\
058472&01 04 09.52&$-$72 50 15.3&13.34& 1.82& 0.96& 3.520& -8.02& 181.8&SMC &K0 I   &\nodata        \\
058738&01 04 15.46&$-$72 45 19.9&12.75& 1.60& 0.76& 3.568& -7.61& 162.1&SMC &K2 I   & \nodata       \\
058839&01 04 17.71&$-$71 57 32.5&13.21& 1.81& 0.95& 3.568& -7.15& 192.5&SMC &K2 I   &\nodata        \\
059426&01 04 30.26&$-$72 04 36.1&13.08& 1.80& 0.99& 3.538& -7.80& 167.0&SMC &K5-7 I &K5-M0  \\
059803&01 04 38.16&$-$72 01 27.2&11.98& 1.95& 0.98& 3.512& -9.65& 200.0&SMC &M0-1 I & \nodata       \\
060447&01 04 53.05&$-$72 47 48.5&13.09& 1.64& 0.94& 3.518& -8.32& 172.6&SMC &M0 I   & \nodata       \\
061296&01 05 11.50&$-$72 02 27.5&13.07& 1.77& 0.92& 3.531& -7.97& 160.1&SMC &K7 I   & \nodata       \\
062427&01 05 40.04&$-$71 58 46.4&13.06& 1.73& 0.85& 3.544& -7.68& 170.5&SMC &K5 I   & \nodata       \\
062763&01 05 50.26&$-$71 58 02.2&12.12& 1.20& 0.72&  \nodata &  \nodata &  46.3&Fgd &K V    & \nodata       \\
063114&01 06 01.37&$-$72 52 43.2&12.83& 1.88& 0.94& 3.538& -8.05& 194.8&SMC &K5-7 I & \nodata       \\
063131&01 06 01.72&$-$72 24 03.8&12.97& 1.67& 0.85& 3.544& -7.77& 168.1&SMC &K5 I   &       \\
063188&01 06 03.21&$-$72 52 16.0&13.07& 1.85& 0.92& 3.568& -7.29& 175.3&SMC &K2 I   & \nodata       \\
064448&01 06 40.21&$-$72 28 45.2&12.68& 1.55& 0.76& 3.544& -8.06& 155.5&SMC &K2-7 I &M0 Ia- \\
064663&01 06 47.62&$-$72 16 11.9&11.87& 1.40& 0.88& 3.531& -9.17& 139.3&SMC &K7 I   & \nodata       \\
066066&01 07 29.36&$-$72 30 45.7&12.57& 1.69& 0.82& 3.544& -8.17& 148.6&SMC &K5 I   & \nodata       \\
066510&01 07 43.12&$-$72 12 15.1&11.69& 1.21& 0.66&  \nodata &  \nodata & -60.1&Fgd &K5 V   & \nodata       \\
066694&01 07 48.88&$-$72 23 42.4&12.52& 1.76& 0.91& 3.544& -8.22& 137.6&SMC &K5 I   & \nodata       \\
066754&01 07 50.91&$-$72 10 46.8&13.02& 1.35& 0.94&  \nodata &  \nodata & -40.1&Fgd &G V    & \nodata       \\
067509&01 08 13.34&$-$72 00 02.9&12.74& 1.68& 0.86& 3.568& -7.62& 153.2&SMC &K2 I   &\nodata        \\
067554&01 08 14.65&$-$72 46 40.8&12.64& 1.62& 0.84& 3.538& -8.24& 195.9&SMC &K5-7 I & \nodata       \\
068648&01 08 52.08&$-$72 23 07.0&12.33& 1.76& 0.87& 3.568& -8.03& 181.4&SMC &K2 I   & \nodata       \\
069317&01 09 17.09&$-$72 12 42.6&12.86& 1.59& 1.08&  \nodata &  \nodata &  13.1&Fgd &M3-4 V &\nodata        \\
070859&01 10 19.89&$-$72 03 34.8&11.06& 1.39& 0.88&  \nodata &  \nodata &  27.2&Fgd &Dwarf  & \nodata       \\
071507&01 10 50.25&$-$72 00 14.5&13.26& 1.74& 0.89& 3.552& -7.29& 152.7&SMC &K3-5 I & \nodata  \\
071566&01 10 53.51&$-$72 25 40.0&13.00& 1.76& 0.89& 3.531& -8.04& 188.8&SMC &K7 I   &  \nodata      \\
081668&01 24 54.03&$-$73 26 49.2&13.14& 1.77& 0.90& 3.544& -7.60& 161.3&SMC & \nodata       & \nodata       \\
081961&01 25 38.80&$-$73 21 55.6&11.84& 1.90& 0.94& 3.528& -9.28& 160.2&SMC & \nodata       & \nodata       \\
082159&01 26 09.91&$-$73 23 15.4&12.71& 1.71& 0.83& 3.573& -7.60& 167.2&SMC & \nodata       &  \nodata      \\
083202&01 29 18.52&$-$73 01 59.3&11.53& 1.08& 0.82& 3.577& -8.73& 158.6&SMC &  \nodata      & \nodata       \\
083593&01 30 33.92&$-$73 18 41.9&12.64& 1.87& 1.00& 3.491& -9.80& 180.1&SMC & \nodata       &M2 Ia  \\
084202&01 33 08.98&$-$73 25 32.5&12.00& 1.33& 0.70&  \nodata &  \nodata &  95.3&Fgd?& \nodata       & \nodata       \\
084392&01 34 08.70&$-$73 06 04.5&11.48& 1.55& 1.33&  \nodata &  \nodata & -21.4&Fgd &  \nodata      & \nodata       \\
\enddata
\tablenotetext{a}{Star ID, coordinates, and photometry are from Massey 2002a.}
\tablenotetext{b}{Based upon spectral type, if available, or $V-R$ if not. See text.}
\tablenotetext{c}{Radial velocity in units of km~s$^{-1}$.}
\tablenotetext{d}{Literature spectral types are from Elias, Frogel \& Humphreys 1985.}
\end{deluxetable}
\begin{deluxetable}{l c c c c c c r r l l l}
\tabletypesize{\footnotesize}
\tablewidth{0pc}
\tablenum{2}
\tablecolumns{12}
\tablecaption{\label{tab:LMC}Red Stars Seen Towards the LMC\tablenotemark{a}}
\tablehead{
\multicolumn{10}{c}{}
&\multicolumn{2}{c}{Spectral Type} \\ \cline{11-12}
\colhead{Star}
& \colhead{$\alpha_{2000}$}
& \colhead{$\delta_{2000}$}
& \colhead{V}
& \colhead{B-V}
& \colhead {V-R}
& \colhead {log $T_{\rm eff}$\tablenotemark{b}}
& \colhead {$M_{\rm bol}$\tablenotemark{b}}
& \colhead {RV\tablenotemark{c}}
& \colhead {Member?}
& \colhead {New}
& \colhead {Lit.\tablenotemark{d}}
}
\startdata
009392&04 50 58.71&$-$69 14 03.2&12.88& 1.95& 1.07& 3.533& -7.99& 263.9&LMC &\nodata        & \nodata       \\
010895&04 51 30.99&$-$69 14 52.0&13.55& 1.94& 1.23& 3.479& -9.31& 263.2&LMC & \nodata       &\nodata        \\
011656&04 51 47.29&$-$69 19 25.1&13.01& 1.89& 1.01& 3.553& -7.39& 276.9&LMC & \nodata       &  \nodata      \\
016332&04 53 14.78&$-$69 12 18.3&13.73& 2.17& 1.39& 3.477& -9.22& 276.8&LMC & \nodata       & \nodata       \\
016554&04 53 18.50&$-$69 17 03.5&12.88& 1.20& 0.97& 3.566& -7.37& 267.8&LMC & \nodata       &  \nodata      \\
017261&04 53 30.84&$-$69 17 49.8&13.00& 1.03& 1.20& 3.489& -9.39& 262.1&LMC & \nodata       & \nodata       \\
017338&04 53 32.34&$-$69 01 17.8&10.93& 1.54& 0.85&  \nodata &  \nodata & -13.9&Fgd & \nodata       & \nodata       \\
018456&04 53 51.19&$-$68 50 04.3&12.57& 1.60& 0.98&  \nodata &  \nodata &  83.0&Fgd & \nodata       &\nodata        \\
021369&04 54 36.90&$-$69 20 22.7&11.26& 1.77& 0.87& 3.600& -8.63& 256.1&LMC &\nodata        &\nodata        \\
021480&04 54 38.56&$-$69 11 17.4&13.19& 1.82& 1.48& 3.477& -9.76& 256.1&LMC & \nodata       &\nodata        \\
021534&04 54 39.46&$-$69 04 36.7&12.63& 2.02& 1.07& 3.533& -8.24& 270.4&LMC &\nodata        &\nodata        \\
022204&04 54 49.77&$-$69 30 03.0&12.72& 1.98& 0.99& 3.560& -7.60& 261.7&LMC & \nodata       &\nodata        \\
023095&04 55 03.09&$-$69 29 13.4&14.38& 2.00& 1.79& 3.477& -8.57& 252.0&LMC &\nodata        &\nodata        \\
024014&04 55 16.11&$-$69 19 12.8&12.82& 1.47& 1.23& 3.479&-10.04& 246.5&LMC &\nodata        &\nodata        \\
024410&04 55 21.72&$-$69 47 17.2&14.45& 2.07& 1.57& 3.477& -8.50& 257.1&LMC & \nodata       &\nodata        \\
024987&04 55 30.05&$-$69 29 11.1&12.08& 2.04& 1.09& 3.526& -8.97& 260.7&LMC &\nodata        & \nodata       \\
025818&04 55 41.86&$-$69 26 24.8&11.72& 2.06& 1.05& 3.539& -8.99& 253.6&LMC &\nodata        & \nodata       \\
026286&04 55 48.28&$-$69 24 07.1&12.37& 1.96& 1.04& 3.543& -8.27& 256.1&LMC & \nodata       & \nodata       \\
028780&04 56 23.70&$-$69 42 11.9&12.76& 1.83& 0.96& 3.570& -7.45& 262.8&LMC & \nodata       & \nodata       \\
029153&04 56 28.30&$-$69 40 37.6&12.85& 1.79& 0.98& 3.563& -7.44& 268.6&LMC &\nodata        &\nodata        \\
030861&04 56 48.61&$-$69 39 55.9&12.25& 1.92& 1.09& 3.526& -8.80& 249.9&LMC & \nodata       & \nodata       \\
030929&04 56 49.63&$-$69 48 32.0&12.06& 1.64& 0.73& 3.647& -7.43& 248.8&LMC &\nodata        &\nodata        \\
035415&04 57 44.66&$-$69 30 35.0&13.37& 1.93& 1.10& 3.523& -7.78& 260.9&LMC & \nodata       &\nodata        \\
038347&04 58 21.08&$-$69 33 38.3&11.30& 1.35& 0.73&  \nodata &  \nodata &   5.1&Fgd & \nodata       & \nodata       \\
054365&05 02 09.57&$-$70 25 02.4&13.26& 1.85& 1.10& 3.506& -8.45& 237.0&LMC &M3 I   &\nodata        \\
058820&05 03 15.36&$-$70 17 41.9&13.25& 1.81& 1.03& 3.544& -7.37& 251.8&LMC &M0 I   & \nodata       \\
062090&05 04 05.10&$-$70 22 46.7&12.50& 1.96& 1.00& 3.531& -8.40& 243.8&LMC &M1 I   & \nodata       \\
062353&05 04 09.92&$-$70 12 18.0&12.86& 1.49& 1.00& 3.531& -8.04& 237.2&LMC &M1 I   & \nodata       \\
064048&05 04 41.79&$-$70 42 37.2&13.28& 1.89& 1.19& 3.506& -8.43& 240.4&LMC &M3 I   & \nodata       \\
064706&05 04 54.13&$-$70 33 18.9&12.79& 1.63& 0.98& 3.538& -7.96& 238.3&LMC &M0-1 I & \nodata       \\
065558&05 05 10.03&$-$70 40 03.2&12.62& 1.89& 1.01& 3.544& -8.00& 246.3&LMC &M0 I   & \nodata       \\
066778&05 05 33.44&$-$70 33 47.1&12.92& 1.40& 1.19& 3.484& -9.70& 240.6&LMC &M4 I   &\nodata        \\
067982&05 05 56.61&$-$70 35 24.0&12.76& 1.93& 1.09& 3.477&-10.19& 244.1&LMC &M4.5 I & \nodata       \\
068098&05 05 58.92&$-$70 29 14.6&13.11& 1.90& 1.04& 3.531& -7.79& 240.2&LMC &M1 I   & \nodata       \\
068125&05 05 59.56&$-$70 48 11.4&13.43& 1.83& 1.20& 3.484& -9.19& 224.3&LMC &M4 I   &M5Iab  \\
069960&05 06 36.42&$-$70 32 38.7&13.10& 1.91& 1.02& 3.518& -8.18& 242.3&LMC &M2 I   & \nodata       \\
071357&05 07 05.62&$-$70 32 44.3&11.70& 2.07& 1.09& 3.531& -9.20& 241.8&LMC &M1 I   & \nodata       \\
072727&05 07 32.52&$-$70 39 04.6&13.08& 2.15& 1.20& 3.518& -8.20& 234.4&LMC &M2 I   &\nodata        \\
106201&05 17 09.11&$-$69 32 21.1&13.29& 1.51& 1.24& 3.518& -7.99& 261.0&LMC &M2 I   & \nodata       \\
109106&05 17 56.51&$-$69 40 25.4&12.96& 1.85& 1.02& 3.518& -8.32& 248.8&LMC &M2 I   &\nodata        \\
113364&05 19 03.35&$-$69 39 55.2&11.70& 1.46& 0.93& 3.531& -9.20& 253.0&LMC &M1 I   & \nodata       \\
116895&05 19 53.34&$-$69 27 33.4&12.43& 1.92& 1.03& 3.506& -9.28& 264.3&LMC &M3 I   &\nodata        \\
119219&05 20 23.69&$-$69 33 27.3&12.14& 2.04& 0.98& 3.506& -9.57& 259.4&LMC &M3 I   & \nodata       \\
123778&05 21 28.06&$-$69 30 16.5&13.49& 1.78& 1.10& 3.506& -8.22& 274.2&LMC &M3 I   & \nodata       \\
124836&05 21 43.54&$-$69 21 27.6&13.19& 1.59& 1.01& 3.553& -7.21& 274.6&LMC & \nodata       &\nodata        \\
126683&05 22 11.01&$-$69 17 24.2&11.60& 1.24& 0.67&  \nodata &  \nodata &  93.3&Fgd &K2 V   &  \nodata      \\
128130&05 22 31.21&$-$69 34 05.1&13.07& 1.86& 1.00& 3.518& -8.21& 259.0&LMC &M2 I   & \nodata       \\
130426&05 23 02.84&$-$69 20 37.1&13.18& 1.88& 1.05& 3.531& -7.72& 256.4&LMC &M1 I   & \nodata       \\
131735&05 23 34.09&$-$69 19 07.0&12.65& 1.84& 0.89& 3.556& -7.71& 234.8&LMC &K7 I   & \nodata       \\
134383&05 25 44.95&$-$69 04 48.9&13.46& 1.65& 1.21& 3.506& -8.25& 268.4&LMC &M3 I   &M3 I   \\
135720&05 26 27.52&$-$69 10 55.5&13.57& 1.85& 1.35& 3.506& -8.14& 269.9&LMC &M3 I   & \nodata       \\
135754&05 26 28.32&$-$69 07 57.4&13.07& 1.96& 1.05& 3.531& -7.83& 279.0&LMC &M1 I   & \nodata       \\
136042&05 26 34.92&$-$68 51 40.1&12.24& 1.08& 1.09& 3.531& -8.66& 266.0&LMC &M1 I   &M2 I + \\
136348&05 26 42.20&$-$68 56 38.7&13.11& 1.89& 1.05& 3.531& -7.79& 276.7&LMC &M1 I   &  \nodata      \\
136378&05 26 42.79&$-$68 57 13.4&13.28& 1.97& 1.11& 3.506& -8.43& 301.0&LMC &M3 I   & \nodata       \\
137624&05 27 10.38&$-$69 16 17.6&13.16& 1.88& 1.02& 3.544& -7.46& 279.1&LMC &M0 I   & \nodata       \\
137818&05 27 14.33&$-$69 11 10.7&13.33& 1.74& 1.20& 3.506& -8.38& 274.9&LMC &M3 I   & \nodata       \\
138405&05 27 26.86&$-$69 00 02.0&13.08& 1.83& 1.02& 3.544& -7.54& 271.8&LMC &M0 I   &M0 Iab \\
138475&05 27 28.16&$-$69 00 36.0&12.65& 1.66& 1.03& 3.544& -7.97& 271.0&LMC &M0 I   &M1 Ia- \\
138552&05 27 29.84&$-$67 14 12.9&12.80& 1.54& 1.18& 3.531& -8.10& 296.1&LMC & \nodata       &M1 Ia  \\
139027&05 27 39.72&$-$69 09 01.1&12.13& 1.15& 0.92& 3.556& -8.23& 281.4&LMC &K7 I   &M1 Ia  \\
139413&05 27 47.62&$-$69 13 20.3&12.68& 1.53& 1.17& 3.506& -9.03& 272.8&LMC &M3 I   & \nodata       \\
139588&05 27 51.22&$-$67 18 04.3&13.19& 1.83& 1.05& 3.531& -7.71& 292.9&LMC &M1 I   & \nodata       \\
139591&05 27 51.28&$-$69 10 45.8&12.54& 1.40& 0.96& 3.525& -8.53& 265.5&LMC &M1-2 I & \nodata       \\
140006&05 28 00.12&$-$69 07 42.3&13.05& 1.71& 0.97& 3.512& -8.44& 267.8&LMC &M2-3 I &M0 Ia  \\
140296&05 28 06.11&$-$69 07 13.5&13.12& 1.87& 1.18& 3.525& -7.95& 271.2&LMC &M1-2 I &M0 Ia  \\
140403&05 28 08.18&$-$69 13 10.8&13.01& 2.01& 1.15& 3.484& -9.61& 267.7&LMC &M3-5 I &\nodata        \\
140782&05 28 16.01&$-$69 12 01.1&13.00& 1.67& 1.03& 3.531& -7.90& 271.4&LMC &M1 I   & \nodata       \\
140912&05 28 18.69&$-$69 07 34.7&12.83& 1.13& 0.97& 3.531& -8.07& 276.4&LMC &M1 I   &M1 Ia- \\
141377&05 28 28.01&$-$69 12 57.2&10.93& 1.61& 0.70& 3.657& -8.48& 272.1&LMC &K0 I   & \nodata       \\
141507&05 28 30.42&$-$69 00 44.7&12.98& 1.90& 1.01& 3.544& -7.64& 285.1&LMC &M0 I   & \nodata       \\
141568&05 28 31.63&$-$69 05 31.2&13.23& 2.02& 1.19& 3.512& -8.26& 272.5&LMC &M2-3 I &M2 Iab \\
142102&05 28 43.26&$-$67 18 28.5&12.84& 1.89& 0.99& 3.556& -7.52& 311.8&LMC &K7 I   & \nodata       \\
142202&05 28 45.59&$-$68 58 02.3&12.15& 1.65& 1.03& 3.538& -8.60& 272.3&LMC &M0-M1 I&M0-M1  \\
142907&05 29 00.86&$-$68 46 33.6&13.05& 1.89& 1.06& 3.531& -7.85& 273.1&LMC &M1 I   & \nodata       \\
143035&05 29 03.58&$-$69 06 46.3&13.52& 1.93& 1.27& 3.484& -9.10& 268.3&LMC &M3-4.5 & \nodata       \\
143137&05 29 05.59&$-$67 18 18.0&12.79& 1.11& 0.91& 3.544& -7.83& 314.5&LMC &M0 I   &M0 Iab \\
143280&05 29 08.49&$-$69 12 18.6&13.27& 1.94& 1.13& 3.506& -8.44& 269.3&LMC &M3 I   & \nodata       \\
143877&05 29 21.10&$-$68 47 31.5&11.82& 1.94& 0.95& 3.556& -8.54& 273.2&LMC &K7 I   &M1Ia   \\
143898&05 29 21.49&$-$69 00 20.3&11.96& 0.55& 0.76& 3.525& -9.11& 285.5&LMC &M1-2 I & \nodata       \\
144217&05 29 27.66&$-$69 08 50.3&12.23& 1.67& 1.13& 3.531& -8.67& 267.9&LMC &M1 I   &M1 Ia  \\
145013&05 29 42.32&$-$68 57 17.3&12.15& 1.89& 1.16& 3.518& -9.13& 273.0&LMC &M2 I   &M1Ia   \\
145112&05 29 44.02&$-$69 05 50.2&12.31& 2.08& 1.07& 3.512& -9.18& 263.5&LMC &M2-3   &\nodata        \\
145716&05 29 54.85&$-$69 04 15.6&12.49& 1.86& 0.96& 3.538& -8.26& 285.0&LMC &M0-1   &\nodata        \\
145728&05 29 55.04&$-$67 18 36.9&12.45& 1.19& 1.02& 3.506& -9.26& 308.2&LMC &M3 I   &M1Ia + \\
146126&05 30 02.36&$-$67 02 45.0&11.17& 1.80& 0.84& 3.568& -9.05& 314.3&LMC &K5 I   & \nodata       \\
146244&05 30 04.63&$-$68 47 28.9&12.92& 1.92& 0.98& 3.556& -7.44& 275.0&LMC &K7 I   &M0 Iab \\
146266&05 30 04.99&$-$69 03 59.9&13.15& 1.84& 1.01& 3.556& -7.21& 269.2&LMC &K7 I   & \nodata       \\
146548&05 30 09.67&$-$69 11 03.9&13.80& 2.08& 1.18& 3.550& -6.70& 277.2&LMC &K7-M0 I&\nodata        \\
147199&05 30 21.00&$-$67 20 05.7&12.73& 1.57& 1.20& 3.484& -9.89& 309.4&LMC &M4 I   &M1 Ia  \\
147257&05 30 22.20&$-$67 06 31.4&12.76& 1.45& 0.94& 3.531& -8.14& 302.1&LMC &M1 I   & \nodata       \\
147276&05 30 22.49&$-$67 05 05.9&11.94& 1.33& 0.71&  \nodata &  \nodata &  63.3&Fgd &K2 V   & \nodata       \\
147372&05 30 24.36&$-$67 29 13.0&11.91& 1.31& 0.73&  \nodata &  \nodata &  56.1&Fgd &K2 V   & \nodata       \\
147479&05 30 26.37&$-$69 30 24.7&12.78& 1.90& 1.02& 3.506& -8.93& 266.6&LMC &M3 I   & \nodata       \\
147928&05 30 33.55&$-$67 17 15.4&12.38& 1.30& 0.95& 3.556& -7.98& 291.1&LMC &K7 I   &M2 I + \\
148035&05 30 35.61&$-$68 59 23.6&13.88& 1.66& 1.38& 3.484& -8.74& 284.5&LMC &M4 I   & \nodata       \\
148041&05 30 35.69&$-$67 12 04.3&13.06& 1.81& 0.99& 3.531& -7.84& 308.2&LMC &M1 I   & \nodata       \\
148381&05 30 41.58&$-$69 15 33.7&12.24& 1.86& 1.10& 3.477&-10.71& 272.6&LMC &M4.5-5 & \nodata       \\
148409&05 30 42.10&$-$69 05 23.2&13.32& 1.81& 1.05& 3.538& -7.43& 268.9&LMC &M0-1 I &M1 Iab \\
148600&05 30 45.25&$-$67 07 59.2&13.23& 1.90& 1.11& 3.506& -8.48& 305.2&LMC &M3 I   &\nodata        \\
149026&05 30 52.38&$-$67 17 34.5&12.80& 1.43& 0.93& 3.531& -8.10& 307.6&LMC &M1 I   &M2 I + \\
149065&05 30 53.17&$-$67 30 52.0&12.09& 1.21& 0.68&  \nodata &  \nodata &  -7.9&Fgd &K0 V   & \nodata       \\
149560&05 31 00.62&$-$69 10 39.6&13.05& 0.60& 1.20& 3.489& -9.34& 279.9&LMC &Comp  I& \nodata       \\
149587&05 31 01.19&$-$69 10 59.2&12.51& 0.69& 0.82& 3.544& -8.11& 278.1&LMC &M0 I   &\nodata        \\
149721&05 31 03.50&$-$69 05 40.0&12.71& 1.86& 0.97& 3.562& -7.58& 277.3&LMC &K5-7 I &M1 Iab \\
149767&05 31 04.33&$-$69 19 02.9&13.10& 2.03& 1.29& 3.506& -8.61& 274.4&LMC &M3 I   & \nodata       \\
150040&05 31 09.35&$-$67 25 55.1&12.81& 1.96& 1.20& 3.484& -9.81& 280.0&LMC &M4 I   &M4 Ia- \\
150396&05 31 15.58&$-$69 03 58.8&13.26& 1.81& 1.15& 3.525& -7.81& 271.5&LMC &M1-2 I & \nodata       \\
150577&05 31 18.56&$-$69 09 28.2&13.27& 1.80& 1.08& 3.562& -7.02& 277.8&LMC &K5-7 I & \nodata       \\
150976&05 31 25.82&$-$69 21 17.9&13.17& 1.85& 1.06& 3.531& -7.73& 275.9&LMC &M1 I   & \nodata       \\
152132&05 31 47.50&$-$67 23 03.3&13.16& 1.90& 1.06& 3.544& -7.46& 292.5&LMC &M0 I   &M0 Ia- \\
153298&05 32 08.91&$-$67 11 18.6&13.11& 1.85& 1.03& 3.531& -7.79& 300.9&LMC &M1 I   & \nodata       \\
153866&05 32 19.30&$-$67 25 00.5&13.16& 1.82& 1.01& 3.531& -7.74& 297.2&LMC &M1 I   & \nodata       \\
154311&05 32 27.54&$-$69 16 53.0&12.56& 1.89& 1.06& 3.506& -9.15& 261.0&LMC &M3 I   &\nodata        \\
154542&05 32 31.52&$-$69 20 25.7&13.02& 1.94& 1.01& 3.544& -7.60& 272.6&LMC &M0 I   &\nodata        \\
154729&05 32 35.44&$-$69 07 51.9&13.21& 1.51& 1.08& 3.525& -7.86& 292.0&LMC &M1-2 I &\nodata        \\
155529&05 32 50.32&$-$67 27 45.3&13.34& 1.84& 1.20& 3.506& -8.37& 292.3&LMC &M3 I   &\nodata        \\
156794&05 33 14.53&$-$67 03 48.5&12.95& 1.79& 1.04& 3.531& -7.95& 302.7&LMC &M1 I   &\nodata        \\
157401&05 33 26.88&$-$67 04 13.7&12.27& 1.99& 1.06& 3.506& -9.44& 299.1&LMC &M3 I   & \nodata       \\
157533&05 33 29.67&$-$67 31 38.0&13.16& 1.50& 0.99& 3.568& -7.06& 302.2&LMC &K5 I   &M1 Ia  \\
158317&05 33 44.60&$-$67 24 16.9&13.35& 1.96& 1.12& 3.518& -7.93& 301.2&LMC &M2 I   & \nodata       \\
158646&05 33 52.26&$-$69 11 13.2&13.10& 2.23& 1.33& 3.496& -8.97& 288.3&LMC &M3-4 I & \nodata       \\
159893&05 34 19.57&$-$68 59 36.4&13.10& 1.96& 1.06& 3.536& -7.69& 289.9&LMC &\nodata        &\nodata        \\
159974&05 34 21.49&$-$69 21 59.8&12.72& 1.77& 0.91& 3.580& -7.38& 250.7&LMC &K2-5 I & \nodata       \\
160170&05 34 25.97&$-$69 21 47.7&11.03& 1.53& 0.82&  \nodata &  \nodata &  42.3&Fgd &K2 V   &\nodata        \\
160518&05 34 33.90&$-$69 15 02.3&13.10& 1.89& 1.17& 3.525& -7.97& 300.7&LMC &M1-2 I & \nodata       \\
161078&05 34 47.07&$-$69 29 00.1&12.91& 1.61& 1.02& 3.544& -7.71& 269.4&LMC &M0 I   & \nodata       \\
162635&05 35 24.61&$-$69 04 03.2&14.23& 2.33& 1.36& 3.531& -6.67& 296.9&LMC &M1 I   &\nodata        \\
163007&05 35 32.84&$-$69 04 18.6&13.07& 1.51& 1.04& 3.556& -7.29& 293.8&LMC &K7 I   &\nodata        \\
163466&05 35 43.86&$-$68 51 21.1&12.45& 1.76& 1.05& 3.539& -8.26& 284.6&LMC & \nodata       & \nodata       \\
163814&05 35 52.01&$-$69 22 28.5&12.75& 1.80& 0.99& 3.544& -7.87& 267.1&LMC &M0 I   & \nodata       \\
164506&05 36 06.44&$-$68 56 40.8&12.87& 1.44& 0.97& 3.566& -7.38& 288.0&LMC & \nodata       &\nodata       \\
164709&05 36 10.56&$-$68 54 40.5&12.02& 0.40& 0.67& 3.667& -7.33& 287.8&LMC & \nodata       &\nodata        \\
165242&05 36 20.42&$-$68 56 18.9&13.97& 1.95& 1.28& 3.477& -8.98& 289.2&LMC &\nodata        &\nodata        \\
165543&05 36 26.91&$-$69 23 50.7&10.98& 1.62& 0.81& 3.620& -8.73& 277.1&LMC &K0 I   & \nodata       \\
166155&05 36 40.60&$-$69 23 16.4&12.94& 1.51& 0.96& 3.556& -7.42& 263.5&LMC &K7 I   & \nodata       \\
168047&05 37 20.65&$-$69 19 38.2&12.47& 1.54& 0.97& 3.568& -7.75& 262.1&LMC &K2-7 I & \nodata       \\
168290&05 37 26.37&$-$68 47 40.1&13.23& 2.03& 1.18& 3.496& -8.87& 305.3&LMC & \nodata       & \nodata       \\
168469&05 37 30.70&$-$69 02 33.2&13.50& 2.24& 1.14& 3.562& -6.79& 265.8&LMC &K5-7 I &\nodata        \\
168757&05 37 36.96&$-$69 29 23.5&14.08& 1.77& 1.34& 3.506& -7.63& 272.3&LMC &M3 I   &\nodata        \\
169049&05 37 43.16&$-$69 24 59.6&12.65& 2.02& 1.14& 3.518& -8.63& 264.5&LMC &M1-3 I &\nodata        \\
169142&05 37 45.15&$-$69 20 48.2&12.11& 0.91& 0.91& 3.496& -9.96& 264.5&LMC &M3-4 I &\nodata        \\
169754&05 37 58.77&$-$69 14 23.7&13.21& 2.15& 1.13& 3.591& -6.77& 275.2&LMC &K2-3 I & \nodata       \\
170079&05 38 06.71&$-$69 17 29.5&14.60& 2.30& 1.60& 3.506& -7.11& 256.5&LMC &M3 I   &\nodata        \\
170452&05 38 16.10&$-$69 10 10.9&13.99& 2.39& 1.50& 3.477& -8.96& 289.4&LMC &M4.5-5 &\nodata        \\
170455&05 38 16.20&$-$69 23 31.7&12.08& 1.36& 0.90&  \nodata &  \nodata & -29.1&Fgd &Dwarf  & \nodata       \\
170539&05 38 18.24&$-$69 17 42.1&13.86& 2.14& 1.29& 3.506& -7.85& 261.1&LMC &M3 I   & \nodata       \\
173854&05 39 46.25&$-$69 19 28.1&13.60& 2.08& 1.19& 3.531& -7.30& 244.8&LMC &M1 I   &\nodata        \\
174324&05 40 07.72&$-$69 20 05.1&13.83& 1.91& 1.26& 3.512& -7.66& 255.3&LMC &M2-3 I & \nodata       \\
174543&05 40 17.13&$-$69 27 53.7&12.97& 1.58& 1.06& 3.506& -8.74& 246.9&LMC &M3 I   &\nodata        \\
174714&05 40 24.48&$-$69 21 16.6&13.13& 1.98& 1.21& 3.477& -9.82& 251.0&LMC &M4-5 I & \nodata       \\
174742&05 40 25.38&$-$69 15 30.2&12.50& 1.63& 0.92& 3.550& -8.00& 251.4&LMC &K7-M0 I& \nodata       \\
175015&05 40 37.04&$-$69 26 20.1&13.31& 1.92& 1.15& 3.506& -8.40& 249.6&LMC &M3 I   & \nodata       \\
175188&05 40 43.80&$-$69 21 57.8&13.52& 1.72& 1.36& 3.512& -7.97& 260.4&LMC &M2-3 I & \nodata       \\
175464&05 40 55.36&$-$69 23 25.0&12.90& 2.20& 1.22& 3.512& -8.59& 245.6&LMC &M2-3 I &\nodata        \\
175549&05 40 59.25&$-$69 18 36.2&13.24& 2.23& 1.39& 3.512& -8.25& 243.9&LMC &M2-3 I &M2 I   \\
175709&05 41 05.17&$-$69 04 42.5&12.74& 1.95& 1.06& 3.544& -7.88& 251.7&LMC &M0 I   &  \nodata      \\
175746&05 41 06.94&$-$69 17 14.8&13.30& 2.06& 1.26& 3.506& -8.41& 262.2&LMC &M3 I   &M1 Ia- \\
176135&05 41 21.89&$-$69 31 48.8&13.06& 2.10& 1.26& 3.506& -8.65& 255.2&LMC &M3 I   & \nodata       \\
176216&05 41 24.60&$-$69 18 12.8&13.66& 1.67& 1.22& 3.556& -6.70& 257.5&LMC &K7 I   &M1 Ia- \\
176335&05 41 29.70&$-$69 27 16.2&12.90& 2.01& 1.03& 3.556& -7.46& 247.4&LMC &K7 I   & \nodata       \\
176695&05 41 43.49&$-$69 28 15.4&12.92& 1.97& 1.03& 3.544& -7.70& 248.9&LMC &M0 I   & \nodata       \\
176715&05 41 44.05&$-$69 12 02.7&13.05& 1.13& 0.98& 3.544& -7.57& 243.7&LMC &M0 I   &M1 I   \\
176890&05 41 50.26&$-$69 21 15.7&12.85& 1.97& 1.01& 3.556& -7.51& 254.8&LMC &K7 I   &M0 Iab \\
177150&05 42 00.84&$-$69 11 37.0&13.80& 1.89& 1.20& 3.531& -7.10& 249.0&LMC &M1 I   &M1 Iab \\
178066&05 42 38.71&$-$69 09 51.4&13.30& 2.00& 1.05& 3.556& -7.06& 245.2&LMC &K7 I   &M2 Ia  \\
178555&05 43 02.16&$-$69 05 49.6&13.04& 1.97& 1.09& 3.544& -7.58& 269.6&LMC &M0 I   & \nodata       \\
\enddata
\tablenotetext{a}{Star ID, coordinates, and photometry are from Massey 2002a.}
\tablenotetext{b}{Based upon spectral type, if available, or $V-R$ if not. See text.}
\tablenotetext{c}{Radial velocity in units of km~s$^{-1}$.}
\tablenotetext{d}{Literature spectral types are from Humphreys
1979.}
\end{deluxetable}
\begin{deluxetable}{l c c c c c }
\tabletypesize{\footnotesize}
\tablewidth{0pc}
\tablenum{3}
\tablecolumns{6}
\tablecaption{\label{tab:teff}Effective Temperatures}
\tablehead{
\colhead{Spectral}
&\multicolumn{4}{c}{Effective Temperatures ($^\circ$K)}
&\colhead{Bolometric} \\ \cline{2-5} 
\colhead{Type}
&\colhead{HM\tablenotemark{a}}
&\colhead{Lee\tablenotemark{b}}
&\colhead{Dyck\tablenotemark{c}}
&\colhead{Adopted}
&\colhead{Corr.~(mag)\tablenotemark{d}} 
}
\startdata
K2~I  &4300& \nodata & 3970 &4000&   -0.97  \\
K5~I  &4000& \nodata & 3520 &3800&   -1.20  \\
K7~I  &3750& \nodata & 3490\tablenotemark{e}&3700&   -1.36  \\
M0~I  &3550&3600 & 3460\tablenotemark{e}&3600&   -1.50  \\
M1~I  &3450&3550 & 3435 &3500&   -1.71 \\
M2~I  &3350&3450 & 3340 &3400&   -2.00 \\
M3~I  &3250&3200 & 3275 &3300&   -2.37 \\
M4~I  &3000&2950 & 3195 &3150&   -3.09 \\
M5~I  &2800&2800 & 3070 &3000&   -4.04 \\
\enddata
\tablenotetext{a}{From Humphreys \& McElroy 1984, Table 2.}
\tablenotetext{b}{From Lee 1970, Table 3.}
\tablenotetext{c}{From the Dyck et al.~1996 effective temperature
scale for red {\it giants}, corrected by $-400^\circ$K.}
\tablenotetext{d}{From the Slesnick et al.\ 2002 relation
between bolometric correction and effective temperature.}
\tablenotetext{e}{Interpolated from spectral types K5 and M1.}
\end{deluxetable}
\begin{deluxetable}{l l c c c c c c c c c c}
\tabletypesize{\footnotesize}
\tablewidth{0pc}
\tablenum{4A}
\tablecolumns{12}
\tablecaption{\label{tab:kurucz}Intrinsic Colors Computed From Kurucz (1992) Model Atmospheres}
\tablehead{
\colhead{}
&\colhead{}
&\multicolumn{2}{c}{Galactic\tablenotemark{a}}
&\colhead{}
&\multicolumn{2}{c}{LMC\tablenotemark{b}}
&\colhead{}
&\multicolumn{2}{c}{SMC\tablenotemark{c}} \\ \cline{3-4} \cline{6-7} \cline{9-10}
\colhead{$T_{\rm eff}$ ($^\circ$K)}
& \colhead{Type\tablenotemark{d}}
& \colhead{$(B-V)_o$}
& \colhead{$(V-R)_o$\tablenotemark{e}}
& \colhead{}
& \colhead{$(B-V)_o$}
& \colhead{$(V-R)_o$\tablenotemark{e}}
& \colhead{}
& \colhead{$(B-V)_o$}
& \colhead{$(V-R)_o$\tablenotemark{e}}
}
\startdata
3500 & M1~I  & 1.79 & 0.92 && 1.82 & 0.92 && 1.84 & 0.99 \\
3750 & K5-7~I& 1.72 & 0.90 && 1.71 & 0.92 && 1.70 & 0.91 \\
4000 & K2~I  & 1.59 & 0.81 && 1.56 & 0.80 && 1.54 & 0.80 \\
\enddata
\tablenotetext{a}{Computed from the Kurucz 1992 Atlas9 models with $\log g=0.0$ and metallicity $\log Z/Z_\odot=0.0$}
\tablenotetext{b}{Computed from the Kurucz 1992 Atlas9 models with $\log g=0.0$ and metallicity $\log Z/Z_\odot=-.3$}
\tablenotetext{c}{Computed from the Kurucz 1992 Atlas9 models with $\log g=0.0$ and  metallicity $\log Z/Z_\odot=-.5$}
\tablenotetext{d}{From Table~\ref{tab:teff} for Galactic stars.}
\tablenotetext{e}{$(V-R)_o$ is on the Cousins system, as described by
Bessel (1983).}
\end{deluxetable}
\begin{deluxetable}{l l c c}
\tabletypesize{\footnotesize}
\tablewidth{0pc}
\tablenum{4B}
\tablecolumns{4}
\tablecaption{\label{tab:bessell}Intrinsic Colors From Bessell et al.~(1989) 
Model Atmospheres}
\tablehead{
\colhead{}
&\colhead{}
&\multicolumn{1}{c}{Galactic\tablenotemark{a}}
&\multicolumn{1}{c}{SMC\tablenotemark{b}} \\
\colhead{$T_{\rm eff}$ ($^\circ$K)}
& \colhead{Type\tablenotemark{c}}
& \colhead{$(V-R)_o$\tablenotemark{d}}
& \colhead{$(V-R)_o$\tablenotemark{d}}
}
\startdata
3000 & M5~I   & 1.95 & 1.69 \\
3200 & M3-4~I & 1.28 & 1.15 \\
3350 & M2-3~I & 0.86 & 0.92 \\
3500 & M1~I   & 0.74 & 0.84 \\
3650 & K7-M0~I& 0.69 & 0.79 \\
3800 & K5~I   & 0.65 & 0.73 \\
\enddata
\tablenotetext{a}{From the Bessell et al.~(1989) 15$\cal M_\odot$ models with
$\log Z/Z_\odot=0$ and $\log g$ varying from -0.11 ($T_{\rm eff}=3800^\circ$)
to -0.52 ($T_{\rm eff}=3000^\circ$).}
\tablenotetext{b}{From the Bessell et al.~(1989) 15$\cal M_\odot$ models with
$\log Z/Z_\odot=-0.5$ and $\log g$ varying from -0.11 ($T_{\rm eff}=3800^\circ$)
to -0.52 ($T_{\rm eff}=3000^\circ$).}
\tablenotetext{c}{From Table~\ref{tab:teff} for Galactic stars.}
\tablenotetext{d}{$(V-R)_o$ is on the Cousins system, as described by
Bessel (1983).}
\end{deluxetable}
\begin{deluxetable}{c c c c c c c c}
\tabletypesize{\footnotesize}
\tablewidth{0pc}
\tablenum{5}
\tablecolumns{8}
\tablecaption{\label{tab:colors}Measured Intrinsic Colors}
\tablehead{
\colhead{}
&\multicolumn{3}{c}{LMC}
&\colhead{}
&\multicolumn{3}{c}{SMC} \\ \cline{2-4} \cline{6-8}
\colhead{Spectral Type}
& \colhead{$(B-V)_o$\tablenotemark{a}}
& \colhead{$(V-R)_o$\tablenotemark{b}}
& \colhead{$N$}
& \colhead{}
& \colhead{$(B-V)_o$\tablenotemark{c}}
& \colhead{$(V-R)_o$\tablenotemark{d}}
& \colhead{$N$}
}
\startdata
K2~I & \nodata          &\nodata & \nodata         &&$1.57\pm0.06$ & $0.83\pm0.04$ &  7\\
K5~I & \nodata          &\nodata & \nodata         &&$1.60\pm0.02$ & $0.84\pm0.01$ & 12\\
K7~I &$1.63\pm0.07$ & $0.92\pm0.01$& 11&&$1.65\pm0.04$ & $0.88\pm0.03$ & 23\\ 
M0~I &$1.61\pm0.08$ & $0.94\pm0.01$& 14&&$1.78\pm0.02$ & $0.94\pm0.01$ & 4 \\
M1~I &$1.66\pm0.04$ & $0.98\pm0.02$& 20&&\nodata           & \nodata& \nodata \\
M2~I &$1.76\pm0.02$ & $1.03\pm0.04$& 5 &&\nodata           & \nodata& \nodata \\
M3~I &$1.75\pm0.03$ & $1.09\pm0.02$& 23&&\nodata           & \nodata& \nodata \\
M4~I &$1.55\pm0.10$ & $1.16\pm0.04$& 5 &&\nodata           & \nodata& \nodata \\
\enddata
\tablenotetext{a}{Corrected by $E(B-V)=0.13$}
\tablenotetext{b}{Corrected by $E(V-R)=0.53\times E(B-V)=0.07$, following
Savage \& Mathis (1973).}
\tablenotetext{c}{Corrected by $E(B-V)=0.06$}
\tablenotetext{b}{Corrected by $E(V-R)=0.53\times E(B-V)=0.03$, following
Savage \& Mathis (1973).}
\end{deluxetable}
\begin{deluxetable}{l c l l c c }
\tabletypesize{\footnotesize}
\tablewidth{0pc}
\tablenum{6A}
\tablecolumns{6}
\tablecaption{\label{tab:smcother}Other Spectroscopically Confirmed RSGs in the SMC}
\tablehead{
\colhead{Id}
&\colhead{V}
&\colhead{Other ID\tablenotemark{a}}
&\colhead{Spectral Type\tablenotemark{b}}
&\colhead{$\log T_{\rm eff}$}
&\colhead{$M_{\rm bol}$}
}
\startdata
\cutinhead{From Massey (2002a)} \\
003196 &13.11& SkKM13  &M1 I           & 3.505&     -8.75  \\                   
018136 &11.98& SkKM63  &M0 Ia          & 3.518&     -9.43  \\                   
021362 &12.89& SkKM78  &K5-M0 I        & 3.531&     -8.15  \\                   
021381 &12.81& SkKM79  &K5-M0 I        & 3.531&     -8.23  \\                   
023401 &12.99& SkKM89  &K5 I           & 3.544&     -7.75  \\                   
035445 &12.74& SkKM144 &M0 Iab         & 3.518&     -8.67  \\                   
069886 &11.74& SkKM319 &M2 Ia          & 3.491&    -10.70  \\                   
\cutinhead{From Elias et al.\ (1985)} \\
101-6  &12.67& SkKM13  &M1 I           & 3.505&     -9.19  \\                   
106-1a &12.24& SkKM63  &M0 Ia          & 3.518&     -9.17  \\                   
105-7  &12.80& SkKM78  &K5-M0 I        & 3.531&     -8.24  \\                   
106-5  &12.95& SkKM79  &K5-M0 I        & 3.531&     -8.09  \\                   
106-7  &13.12& SkKM89  &K5 I           & 3.544&     -7.62  \\                   
106-9  &13.16& SkKM91  &K5-M0 I        & 3.531&     -7.88  \\                   
108-3  &12.56& SkKM110 &M0 I           & 3.518&     -8.85  \\                   
105-11 &12.38& SkKM114 &M0 Iab         & 3.518&     -9.03  \\                   
108-8  &13.19& SkKM129 &K0-2 I         & 3.568&     -7.17  \\                   
105-21 &13.68& SkKM135 &K5-M0 I        & 3.531&     -7.36  \\                   
HV838  &13.35& SkKM142 &M0e I          & 3.518&     -8.06  \\                   
114-3  &12.89& SkKM144 &M0 Iab         & 3.518&     -8.52  \\                   
HV11423&11.77& SkKM205 &M0 Ia          & 3.518&     -9.64  \\                   
115-6  &12.92& SkKM210 &K1-3 Iab       & 3.568&     -7.44  \\                   
116-15 &12.05& SkKM236 &M0 Ia          & 3.518&     -9.36  \\                   
115-17 &13.03& SkKM237 &K5-M0 Iab      & 3.531&     -8.01  \\                   
120-14 &11.96& SkKM275 &K5-M0 Iab      & 3.531&     -9.08  \\                   
HV2084 &12.62& SkKM319 &M2 Ia          & 3.491&     -9.82  \\                   
HV2228 &12.89& SkKM347 &M0 Iab         & 3.518&     -8.52  \\                   
108-2  &12.28& \nodata     &M0 Ia          & 3.518&     -9.13  \\                   
118-18 &13.32& SkKM272?&M0 Ia          & 3.518&     -8.09  \\                   
\enddata
\tablenotetext{a}{Identification from Sanduleak 1989.}
\tablenotetext{b}{Spectral types are all from Elias et al.\ 1985.}
\end{deluxetable}
\begin{deluxetable}{l c l l c c }
\tabletypesize{\footnotesize}
\tablewidth{0pc}
\tablenum{6B}
\tablecolumns{6}
\tablecaption{\label{tab:lmcother}Other Spectroscopically Confirmed RSGs in the LMC}
\tablehead{
\colhead{Id}
&\colhead{V}
&\colhead{Other ID\tablenotemark{a}}
&\colhead{Spectral Type\tablenotemark{b}}
&\colhead{$\log T_{\rm eff}$}
&\colhead{$M_{\rm bol}$}
}
\startdata
\cutinhead{From Massey (2002a)} \\
141430 &12.30& 46-32   &M0 Ia          & 3.544&     -8.32  \\                   
141772 &12.55& 46-34   &M2 Ia          & 3.518&     -8.73  \\                   
156011 &12.11& 53-3    &M0 Ia          & 3.544&     -8.51  \\                   
\cutinhead{From Humphreys (1979)} \\                
46-40  &12.98& \nodata     &M1 Ia          & 3.531&     -7.92  \\                   
45-48  &13.38& \nodata     &M4 Ia-Iab      & 3.484&     -9.24  \\                   
54-35  &12.85& \nodata     &M1 I + B       & 3.531&     -8.05  \\                   
54-47a &13.10& \nodata     &M1 Iab         & 3.531&     -7.80  \\                   
45-2   &12.90& \nodata     &M2 Iab         & 3.518&     -8.38  \\                   
37-32  &12.95& \nodata     &M2 I           & 3.518&     -8.33  \\                   
37-35  &12.89& HV916   &M3 Ia          & 3.506&     -8.82  \\                   
37-24  &13.59& HV2360  &M2 Ia          & 3.518&     -7.69  \\                   
39-33  &12.57& HV888   &M4 Ia          & 3.484&    -10.05  \\                   
46-2   &12.25& HV2450  &M2 Ia          & 3.518&     -9.03  \\                   
61-23  &13.28& \nodata     &M1 Ia-Iab      & 3.531&     -7.62  \\                   
53-3   &12.04& \nodata     &M0 Ia          & 3.544&     -8.58  \\                   
46-31  &13.00& HV2567  &M2 Iab         & 3.518&     -8.28  \\                   
46-51  &12.84& HV2602  &M2 Ia-Iab      & 3.518&     -8.44  \\                   
46-52  &13.40& \nodata     &M1 Iab         & 3.531&     -7.50  \\                   
54-47  &13.02& \nodata     &M0 Iab         & 3.544&     -7.60  \\                   
54-39  &12.76& HV2781  &M1 Ia-Iab      & 3.531&     -8.14  \\                   
54-38  &13.03& \nodata     &M2 Ia          & 3.518&     -8.25  \\                   
54-56  &13.32& \nodata     &M0 Ia-Iab      & 3.544&     -7.30  \\                   
54-44  &13.03& \nodata     &M1 Ia-Iab      & 3.531&     -7.87  \\                   
55-20  &13.09& \nodata     &M2 Ia-Iab      & 3.518&     -8.19  \\                   
52-4   &13.00& HV5914  &M1 Iab         & 3.531&     -7.90  \\                   
\enddata
\tablenotetext{a}{As given in Humphreys 1979.}
\tablenotetext{b}{Spectral types are all from Humphreys 1979.}
\end{deluxetable}

\clearpage

\begin{figure}

\caption{\label{fig:rvhists} Histograms of the radial velocities are shown
for the SMC and the LMC.  The majority of stars have a distribution that
is similar to the radial velocities of the centers of the each galaxy.
The group of lower velocity ($<100$ km~s$^{-1}$) stars
are readily identifiable as foreground red dwarfs.
}
\end{figure}

\begin{figure}

\caption{\label{fig:typehists} Histograms of the spectral types found in
the Milky Way (Elias et al.\ 1985, Table ~20) and the LMC and SMC (from
our Tables 1 and 2, respectively).  There is a progression towards earlier
types.  In the Milky Way the average spectral type is M2~I; in the LMC it is
M1~I, and in the SMC it is K5~I.}
\end{figure}

\begin{figure}

\caption{\label{fig:kurucz} The upper three black curves show Kurucz
(1992) Atlas 9 models corresponding to Galactic metallicity ($\log Z/Z_\odot=0.0$) and low surface gravity ($\log$ g[cgs]=0.0) for
$T_{\rm eff}=4000 ^\circ$K, $3750^\circ$K, and $3500 ^\circ$K.
The spectra below demonstrate that the TiO band strengths predicted
by the Galactic-metallicity $3500^\circ$K are quite similar to what are
observed in M1-~I stars, while the $4000^\circ$K model has lines comparable
to that observed in the K2.5~I standard.
The red and blue curves are $3500 ^\circ$K model computed with
low metallicities (red: $\log Z/Z_\odot=-0.5$, blue: $\log Z/Z_\odot=-1.0$),
which are included  
to show the effects of
low metallicity on the strengths of the TiO bands.  The band strengths in
the low-metallicity models are intermediate between that of the 
higher metallicity 3750$^\circ$K (K5-7~I) and 4000$^\circ$K (K2~I) models,
suggesting that the effect that metallicity has on the appearance on
TiO lines is comparable to that observed in the distribution of spectral types seen in the SMC, LMC, and the Milky Way.
}
\end{figure}
\begin{figure}

\caption{\label{fig:smchrds} The location of the SMC RSGs in the HRD is compared
to three sets of $Z=0.004$ evolutionary tracks: (a) the Geneva models
which include normal mass-loss rates (Charbonnel et al.\ 1993), (b) the
Geneva models which include ``enhanced" ($2\times$ normal) mass-loss rates
(Meynet et al.\ 1994), and (c) the Padova models, 
which also uses normal mass-loss rates  (Fagotto et al.\ 1994).
In none of these cases do the models produce RSGs that are as cool and
luminous as actually observed.  The solid points are been placed in the
diagram using their spectral types to set the effective temperature, while
the open circles have used the photometry to determine the effective
temperature. Red points are data from this paper, while black points are taken
from the literature (i.e., Elias et al.\ 1985).
}
\end{figure}

\begin{figure}

\caption{\label{fig:lmchrds} The location of the LMC RSGs in the HRD is compared
to three sets of $Z=0.008$ evolutionary tracks: (a) the Geneva models
which include normal mass-loss rates (Schaerer et al.\ 1993), (b) the
Geneva models which include ``enhanced" ($2\times$ normal) mass-loss rates
(Meynet et al.\ 1994), and (c) the Padova models, 
which also uses normal mass-loss rates  (Fagotto et al.\ 1994).
In none of these cases do the models produce RSGs that are as cool and
luminous as actually observed.  The solid points are been placed in the
diagram using their spectral types to set the effective temperature, while
the open circles have used the photometry to determine the effective
temperature. Red points are data from this paper, while black points are taken
from the literature (i.e., Humphreys 1979).
}
\end{figure}

\begin{figure}
\caption{\label{fig:oldhrds} Here we show the data from Figs.~\ref{fig:smchrds}
and \ref{fig:lmchrds} plotted as if we had adopted the Galactic effective
temperature scale.  The evolutionary tracks shown are (a) the Geneva
normal mass loss tracks (Z=0.004) 
for the SMC from Charbonnel et al.\ (1993) and (b) the Geneva normal
mass loss tracks (Z=0.008) 
for the LMC from Schaerer et al.\ 1993). Note that there is now
a deficiency of the higher luminosity RSGs in the SMC (a)  compared to that of
the LMC (b). 
Such an effect runs counter to the expectations of stellar evolution,
and gives some addition credence to the corrections adopted earlier.}
\end{figure}

\begin{figure}
\caption{\label{fig:mbolhists} In (a) and (b) we see the relative number of
RSGs as a function of bolometric luminosity if we had made no temperature
correction to the Galactic scale.  The number of high luminosity RSGs drops
fars more steeply in the SMC (a) than in the LMC (b), 
contrary to the expectations
of stellar evolution.  In (c) and (d), we see the same histograms for the
``corrected" temperature scales.  Here the distributions are very similar,
although incompleteness may affect the lowest luminosity bin for the SMC (c).
We have included only the confirmed RSGs from this paper.}
\end{figure}

\end{document}